\documentclass[twocolumn]{aastex62}
\usepackage{amsmath}
\usepackage{multirow}
\usepackage{mathtools}
\usepackage[thinc]{esdiff}
\usepackage{xcolor}

\newcommand{\bz}{$B_\mathrm{LoS}$}

\received{----}
\revised{----}
\accepted{----}
\submitjournal{ApJ}

\shorttitle{frequency dependence of pulse arrival time}
\shortauthors{Das et al.}

\begin{document}

%\title{The magneto-ionic mode transition of electron cyclotron maser emission in hot magnetic stars}
\title{A 3D framework to explore the propagation effects in stars exhibiting electron cyclotron maser emission}

\correspondingauthor{Barnali Das}
\email{barnali@ncra.tifr.res.in}

\author[0000-0001-8704-1822]{Barnali Das}
\affil{National Centre for Radio Astrophysics, Tata Institute of Fundamental Research,  Pune University Campus, Pune-411007, India}

\author[0000-0002-2325-5298]{Surajit Mondal}
\affil{National Centre for Radio Astrophysics, Tata Institute of Fundamental Research,  Pune University Campus, Pune-411007, India}

\author[0000-0002-0844-6563]{Poonam Chandra}
\affil{National Centre for Radio Astrophysics, Tata Institute of Fundamental Research,  Pune University Campus, Pune-411007, India}

%\affiliation{Department of Physics, Royal Military College of Canada, PO Box 17000, Station Forces, Kingston, ON K7K 7B4, Canada}

\begin{abstract}
Recently, coherent radio emission has been discovered from a number of hot magnetic stars, via the process of electron cyclotron maser emission (ECME). This emission, observed in the form of highly circularly polarized pulses, have interesting properties which contain information about the host star. One of the important properties of ECME is the frequency dependence of {\color{black}the} pulse arrival time. This has been attributed to propagation effect by \citet{trigilio2011}, and could explain the sequence observed for CU\,Vir qualitatively \citep{lo2012}. However no quantitative treatment exists for this phenomenon which is a promising tool to estimate {\color{black}the} density in the stellar magnetosphere. Besides, the effect of propagation through the magnetosphere on ECME has been thought to be limited to giving rise to a particular sequence of arrival of pulses, and in some cases producing the upper cut-off frequency for ECME \citep{leto2019}. Here, we present a framework to deal with {\color{black}the} propagation effect by considering continuous refraction in the inner magnetosphere of the star. This framework is capable of incorporating any type of density distribution, and in principle any type of magnetic field, though we limit ourselves to a dipolar magnetic field for this work. We show by simulation that for stars with high obliquity, the propagation effect can influence not only the sequence of arrival of pulses drastically, but also the pulse shapes, and the observability of a pulse from a particular magnetosphere.
\end{abstract}


\section{Introduction} \label{sec:intro}
Coherent radio emission via Electron Cyclotron Maser Emission (ECME) has been observed from a small number of hot magnetic stars \citep[e.g.][]{trigilio2000}, brown dwarfs \citep[e.g.][]{hallinan2006} and planets \citep[e.g.][]{zarka1998}. The observable signatures of this emission include high brightness temperature, high directivity and high degree of circular polarization. 
%Because of the high directivity, the radiation is seen as pulses at certain rotation phases.
%Presence of magnetic field is essential for ECME and the frequency of emission is proportional to the local electron cyclotron frequency. This property has been used to identify the presence of magnetic fields in 
Since presence of magnetic field is a pre-requisite for {\color{black}the} ECME phenomenon, discovery of ECME from cold brown dwarfs  have led to a surprising revelation that such stars can  harbour kilogauss strength magnetic fields \citep{hallinan2006,hallinan2008}.

In case of hot magnetic stars, the magnetic fields are already well measured using spectropolarimetry and Zeeman Doppler Imaging \citep[e.g.][]{kochukhov2014} revealing the dipole like nature of the magnetic field in most cases. 
%Thus ECME is not needed to estimate magnetic fields. 
However, there are a few other properties of ECME which can be exploited to study the properties of the host star. For example, the 
high directivity of the ECME has been used to diagnose the rotation period evolution of the star \citep{trigilio2008,trigilio2011}; and its magneto-ionic mode has been used to estimate the plasma density at the site of emission \citep{leto2019}. 
%As mentioned already in the preceding paragraph, in case of hot magnetic stars, the use of ECME is limited to diagnozing change in stellar rotation period, and inferring plasma density at the site of emission. 
However, one aspect of ECME is yet to be exploited, which is that the high directivity of the phenomenon makes it a sensitive probe to {\color{black}the} local density structures it passes through. 

In a hot massive star, the  interaction between the radiatively driven stellar wind and the magnetic field gives rise to a magnetosphere, which is divided into three parts \citep{andre1988,trigilio2004}: an inner magnetosphere, where the magnetic energy dominates over {\color{black}the} wind kinetic energy and the magnetic field lines are closed;  an outer magnetosphere, where the wind dominates over the magnetic field; and a middle magnetosphere, the transition region between the inner and the outer magnetosphere. Middle magnetosphere is the site of origin of radio emission including {\color{black}the} gyrosynchrotron emission and the ECME.
The boundary of the inner magnetosphere, at which the magnetic field energy equals the wind kinetic energy, is called the Alfv\'en surface. {\color{black}The} inner magnetosphere
(hereafter IM) is the densest part of the magnetosphere with {\color{black}the} largest imprints on the ECME lightcurves.
In this paper, we  present a framework to understand the effect of refraction on the ECME lightcurves while passing through the stellar magnetosphere. This,
in turn, will allow one to acquire information about the density structure in the stellar magnetosphere. 
%This framework is general in the sense that we can use t
This framework is valid for any arbitrary density distribution and, in principle, for any type of arbitrary magnetic field, however, we confine ourselves
to a dipolar magnetic field in this paper.

This paper is structured as follows: in the next section, we present a brief summary about ECME observed from magnetic AB stars (\S\ref{sec:ideal_ecme}). We present our framework in \S\ref{sec:framework}, followed by a few examples of its application in \S\ref{sec:output}. We end this paper with discussion (\S\ref{sec:summary}).

\section{ECME from a star with axi-symmetric dipolar magnetic field}\label{sec:ideal_ecme}

The expected ECME lightcurve from a star with an axi-symmetric dipolar magnetic field consists of two pairs of pulses, each pair consisting of one left circularly polarized (LCP) and one right circularly polarized (RCP) pulses coming from opposite magnetic hemispheres \citep{leto2016}. The LCP and RCP pulses for a given pair lie symmetrically around a magnetic null phase, which is the rotational phase where the line of sight component of the magnetic field is zero ($B_\mathrm{LoS}$). There are two such rotational phases per stellar rotation cycle for a dipolar magnetic field, corresponding to two pairs of ECME pulses. The sequence of arrival of {\color{black}the} RCP and LCP pulses are opposite near the two nulls. Around the null where $B_\mathrm{LoS}$ is changing from positive to negative, the pulse from the northern magnetic hemisphere will arrive first followed by the one from the southern magnetic hemisphere. Around the other null, where $B_\mathrm{LoS}$ is changing from negative to positive, the pulse from the southern magnetic hemisphere will arrive before the one from the northern magnetic hemisphere \citep{leto2016}. This is a consequence of the fact that due to refraction, the pulse from the northern magnetic hemisphere deviates upward, and the one from the south deviates downward \citep{trigilio2011,leto2016}. Note that this picture assumes single refraction at the boundary between {\color{black}the} middle and {\color{black}the} inner magnetosphere at the time of entering the latter.
In reality, we often see a  more complicated sequence of arrival of pulses \citep{das2019a,das2019b}. Moreover, the pulses are almost never seen to lie symmetrically about the magnetic null phases \citep{das2019a,das2019b,leto2019}. We will show subsequently that at least some of these features 
can be explained by the propagation effects alone.

\section{The framework}\label{sec:framework}
Till date, not much work has gone towards understanding the effect of  refraction  experienced by the ECME pulses while travelling through the stellar magnetosphere on its way to the observer.
The importance of refraction was first realized by \citet{trigilio2011} who proposed it to be the cause of the difference in pulse arrival time at two different frequencies in CU\,Vir,
the first known hot magnetic star with ECME \citep{trigilio2000}.
They attributed it to the different amount of deviation suffered due to the refraction in a cold torus with a constant plasma density of $10^9\,\,\mathrm{cm^{-3}}$ \citep[taken from the simulation of][]{leto2006} near the magnetic equator. While doing that, they considered the refraction effect only at the time of entering the cold torus. This scenario was later shown by \citet{lo2012} to be able to correctly reproduce the pulse arrival sequence of the ECME from CU\,Vir at 13 cm and 20 cm, however, the amount of `lag' (difference in rotational phases of arrival for the two frequencies) between {\color{black}the} two pulses could not be reproduced. 
%They assumed a constant plasma density of $10^9$ $\mathrm{cm^{-3}}$ in a torus shaped overdense region inside the IM \citep[also used by][]{trigilio2011} and considered single refraction of radiation at the time of entering the torus.

\begin{figure}
\centering
\includegraphics[trim={5cm 7cm 4cm 1cm},clip,width=0.45\textwidth]{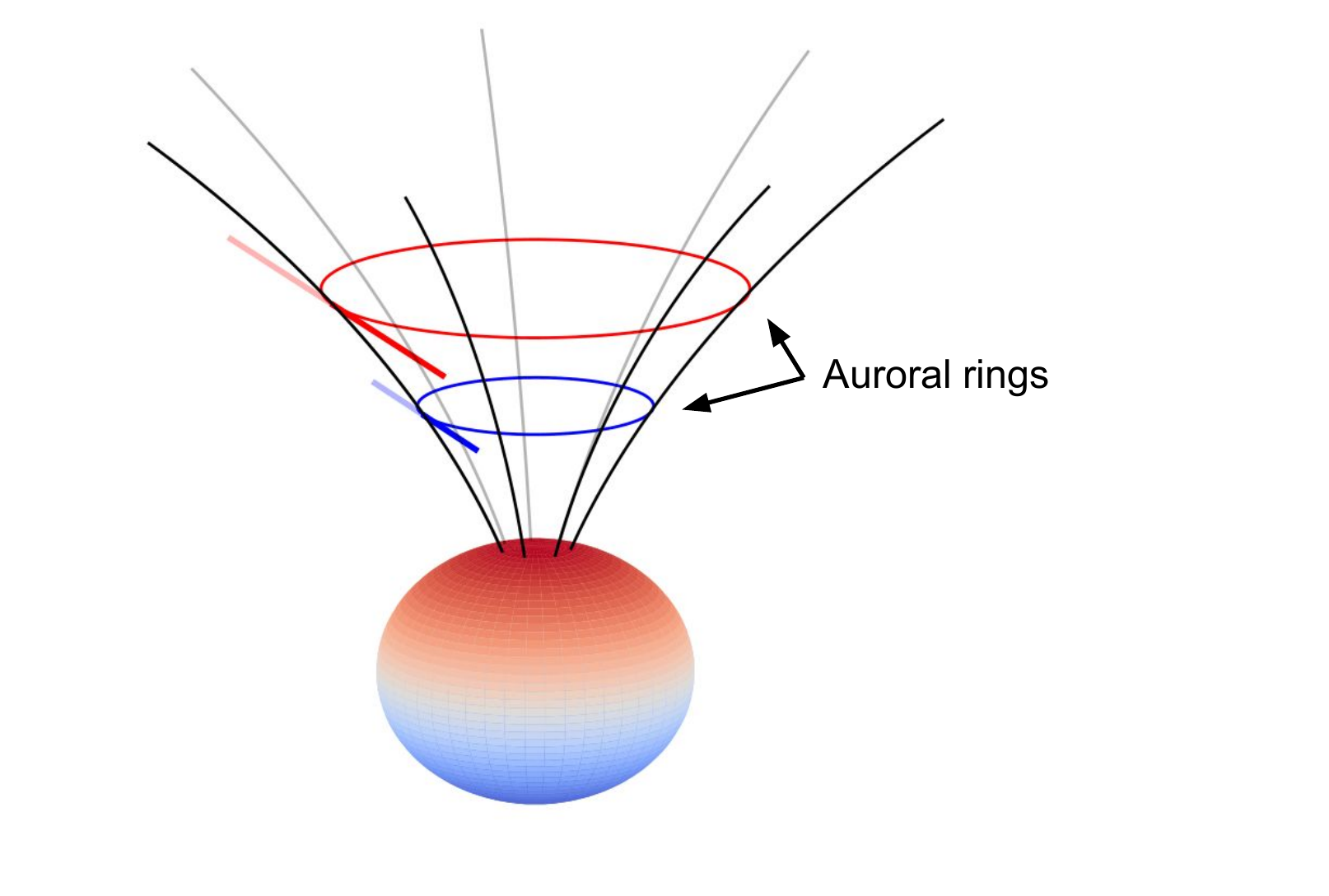}
\caption{The `tangent plane beaming model' for ECME proposed by \citet{trigilio2011}. ECME is produced tangential to the auroral rings so that the direction is perpendicular to the local magnetic field and parallel to the magnetic equatorial plane. Lower frequencies originate farther away from the star and vice-versa.\label{fig:ecme_origin}}
\end{figure}

Here we propose a general framework which will enable one to study the  effect of propagation on the ECME lightcurves for any arbitrary density distribution in the IM.%, as opposed to the `single refraction model' that includes refraction only at the bounding surface of a constant density medium. 
We use the model proposed by \citet{trigilio2011} for the emission of ECME. According to this model, the pulses are emitted tangential to the auroral rings so that they are perpendicular to the local magnetic field vector and parallel to the magnetic equatorial plane (Figure \ref{fig:ecme_origin}).

We divide the whole problem into three parts:
\begin{enumerate}
\item Determine the initial condition, i.e. the ray direction immediately after entering the IM for a given point of origin of the radiation on the auroral circle.
\item Solve the ray path inside the IM.
\item Find out the ray direction after exiting the IM.
\end{enumerate}

We assume that the density in the middle and the outer magnetosphere is low enough so that the refractive indices are unity outside the IM. 
{\color{black} Before presenting our framework in the next section, we would like to clarify that the division of the stellar magnetosphere into three discontinuous parts (inner, middle and outer) is a highly simplified description. In reality, the transition from one region to another is much more complex as demonstrated in various MHD simulatios of hot magnetic stars \citep[e.g.][etc.]{townsend2007, ud-doula2008, ud-doula2013}.}
%In the following subsections, we describe our approach to achieve the goal described in the above three steps.

\subsection{The initial condition}\label{subsec:ini_cond}
\begin{figure}
\centering
\includegraphics[width=0.48\textwidth]{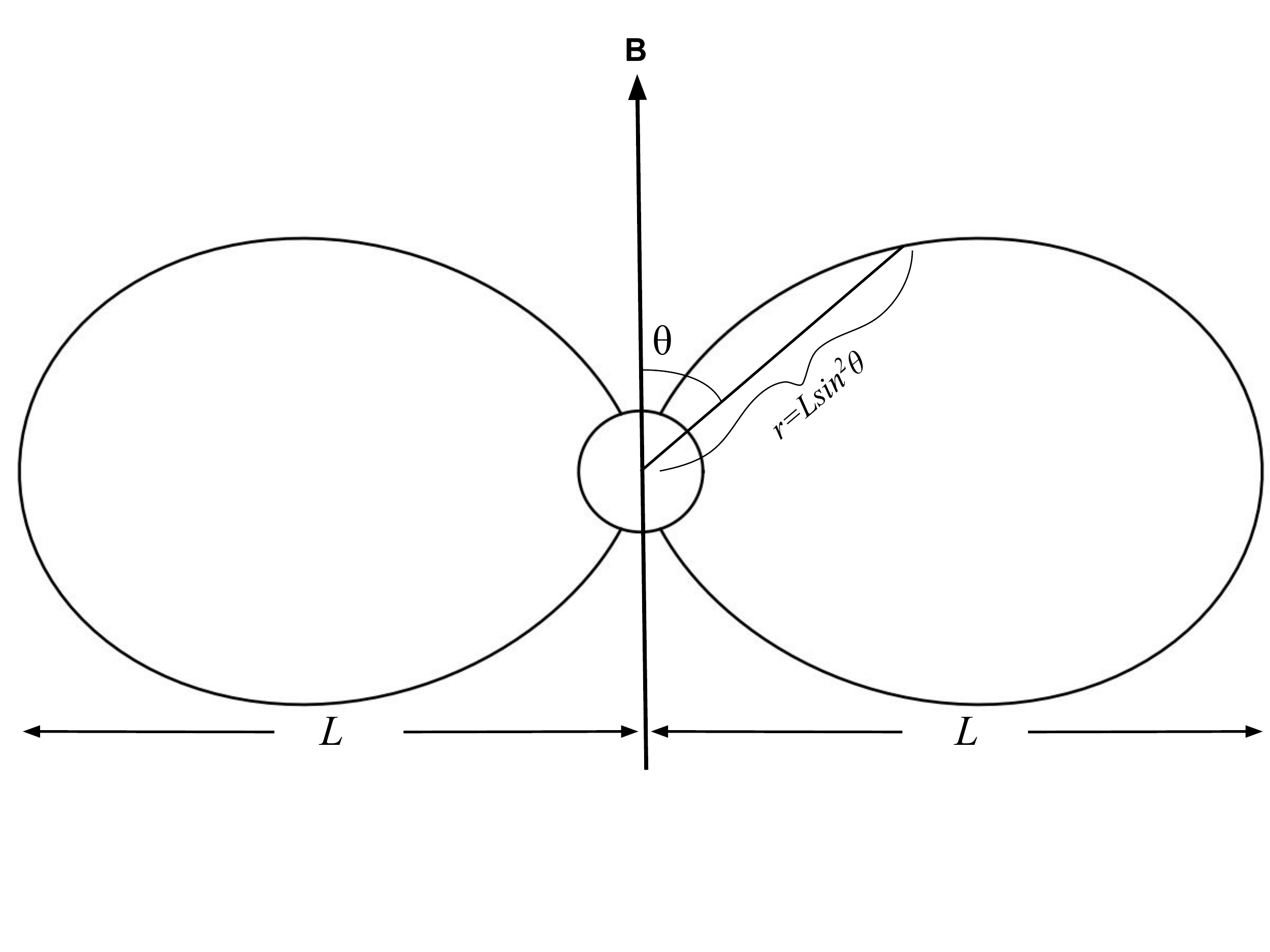}
\caption{The closed magnetic field lines around a star (drawn as the central circle) with a dipolar magnetic field. The axis of the dipole is shown as the vertical arrow and the polar strength on the stellar surface is $B$. The equation of a field line in polar coordinates is $r=L\sin^2\theta$, where $r$ is measured from the centre of the star and $\theta$ is measured from the dipole axis. $L$ is the maximum value of $r$ along that field line, which is obtained for $\theta=90^\circ$, i.e. at the magnetic equator. The IM contains field lines with $L\leq R_\mathrm{A}$.  \label{fig:dipole_line}}
%The field lines lying outside the inner magnetosphere are not closed and have $L>R_\mathrm{A}$.
\end{figure}

In the frame of reference of the magnetic field, we choose the $Z$ axis to lie along the dipole axis. The $X$ and $Y$ axes are arbitrary at this point, but will be defined later (\S\ref{sec:rot_matrix}). The magnetic field lines in polar coordinates are given by $r=L\sin^2\theta$ (Figure \ref{fig:dipole_line}). The IM is bounded by the field line $r=R_\mathrm{A}\sin^2\theta$, where $R_\mathrm{A}$ is the Alfv\'en radius for the star; the field lines inside IM have $L<R_\mathrm{A}$, and those outside have $L>R_\mathrm{A}$.

\begin{figure}
\centering
\includegraphics[trim={7cm 5cm 2cm 0},clip,width=0.45\textwidth]{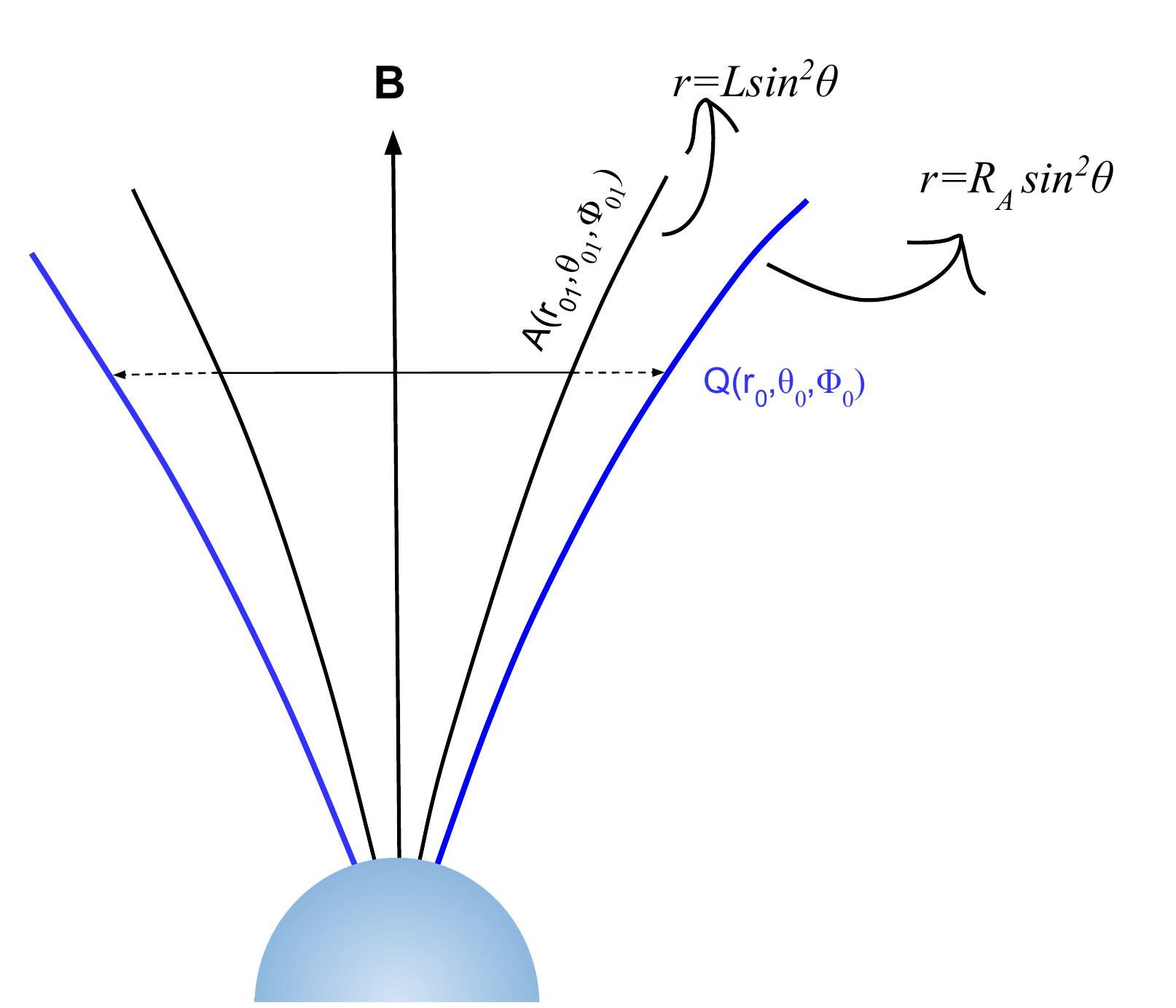}
\caption{A cartoon diagram showing ECME produced at point $A$ that hits the inner magnetosphere boundary at point $Q$.\label{fig:inner_mag_entry}}
\end{figure}

Let the frequency of the ECME be $\nu$, which is related to the local magnetic field strength $B$ as: $\nu\approx s\times2.8B$ ($\nu$ is in MHz and $B$ is in G), where $s$ is the harmonic number. In other words, the ECME at a  frequency $\nu$ will be produced at those points in the middle magnetosphere, at which the magnetic field strength is $B=\nu/(2.8s)$.
For a dipolar magnetic field, these points constitute a (auroral) circle above each magnetic pole (Figure \ref{fig:ecme_origin}) around magnetic field lines given by $r=L\sin^2\theta$. Each point on this auroral circle is a source of the ECME. We consider one such point $A(r_{01},\theta_{01},\phi_{01})$ (Figure \ref{fig:inner_mag_entry}). 
%In that case, the magnetic field strength at point $A$ must be $B=\nu/(2.8s)$.
 If the polar magnetic field on the stellar surface is $B_0$, we have:

%with radius $r_{01}\sin\theta_{01}$, on field lines given by $r_{01}=L\sin^2\theta_{01}$. Note that $L>R_\mathrm{A}$ where $R_\mathrm{A}$ is the Alfv\'en radius for the star. Each point on this auroral circle is a source of ECME. We consider one such point $A(r_{01},\theta_{01},\phi_{01})$ (Figure \ref{fig:inner_mag_entry}). In that case, the magnetic field strength at point $A$ must be $B=\nu/(2.8s)$. If the polar magnetic field on the stellar surface is $B_0$, we have:

%Let a frequency $\nu$ is emitted on a field line characterized by $r=L\sin^2\theta$ at point $A(r_{01},\theta_{01},\phi_{01})$. To obtain the coordinates of $A$, we only need to find $\theta_{01}$ since $r_{01}=L\sin^2\theta_{01}$. To obtain this angle, we note that the magnetic field $B$ at point $A$ is given by $\approx \nu/(2.8s)$, where $s$ is the harmonic number. If the polar magnetic field on the stellar surface is $B_0$, we have:

\begin{align}
&\textbf{B}=\frac{B_0}{r_{01}^3}\left(\cos\theta_{01}\hat{r}+\frac{1}{2}\sin\theta_{01}\hat{\theta}\right)\nonumber\\
\Rightarrow &B=\frac{B_0}{L^3\sin^6\theta_{01}}\sqrt{1-\frac{3}{4}\sin^2\theta_{01}}\nonumber \\
\therefore \quad &  L^3\frac{B}{B_0}-\frac{1}{\sin^6\theta_{01}}\sqrt{1-\frac{3}{4}\sin^2\theta_{01}}=0 \label{eq:finding_ini_theta}
\end{align}

Where we have used $r_{01}=L\sin^2\theta_{01}$. By solving this equation, we will get $\theta_{01}$ and subsequently $r_{01}$.
%=L\sin^2{\theta_{01}}$.

To get the direction of the wave vector ($\textbf{k}$) at the point $A$, immediately after emission, we use the facts that $\textbf{k}$ is perpendicular to both the dipole axis and the local magnetic field $\textbf{B}$ at the point $A(r_{01},\theta_{01},\phi_{01})$. The first condition implies that the $\textbf{k}$ lies in the XY plane. The second condition implies that:

\begin{align}
\textbf{k}\cdot\textbf{B}&=0\nonumber\\
k_xB_x+k_yB_y&=0\nonumber\\
\therefore \frac{k_x}{k_y}&=-\frac{B_{y}}{B_{x}}
\end{align}

This equation gives two possible directions of the wave vector at the given point $A$, which are anti-parallel to each other. 

The next task is to find the point $Q(r_0,\theta_0,\phi_0)$ where the ray will hit the boundary of the IM (Figure \ref{fig:inner_mag_entry}). As we have assumed that the refractive index in the middle magnetosphere is unity, the direction of the wave vector $\textbf{k}$ will not change while travelling from {\color{black}the} points $A$ to $Q$. Since $\textbf{k}$ has no $z$ component, we must have that the $z$ coordinate of {\color{black}the} point $Q$ is same as that of {\color{black}the} point $A$. Thus, we have:
\begin{align*}
(R_A\sin^2\theta_0)\cos\theta_0&=r_{01}\cos\theta_{01}
\end{align*}
By solving this equation, $\theta_0$ can be found. We can then obtain $r_0$ using $r_0=R_A\sin^2\theta_0$.

\begin{figure}
\centering
\includegraphics[trim={2cm 2cm 2cm 2cm}, clip, width=0.45\textwidth]{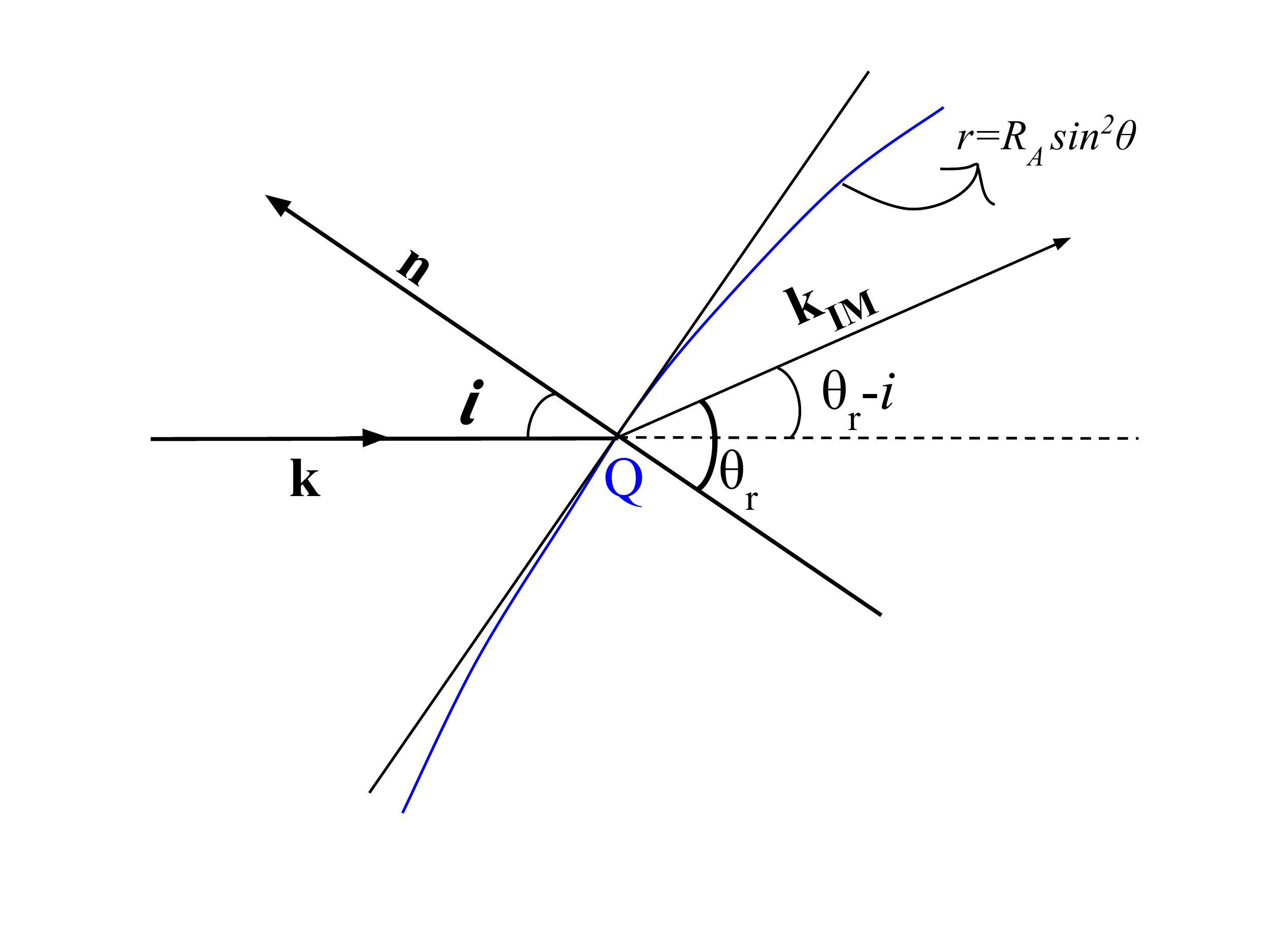}
\caption{The direction of the wave vector $\textbf{k}_\mathrm{IM}$ immediately after entering the inner magnetosphere and its relation with the incident wave vector \textbf{k} and the normal to the surface \textbf{n}. \label{fig:ini_k_IM}}
\end{figure}

To determine  $\phi_0$, we use the fact that the vector connecting the the points $Q$ and $A$ must be parallel to \textbf{k}, i.e.:

\begin{align}
\frac{r_0\sin\theta_0\cos\phi_0-r_{01}\sin\theta_{01}\cos\phi_{01}}{r_0\sin\theta_0\sin\phi_0-r_{01}\sin\theta_{01}\sin\phi_{01}}&=\frac{k_x}{k_y}
\end{align}
%By solving this equation, we get the value of $\phi_0$. 
Now we require the angle of incidence at point $Q$. For that we determine the normal ($\hat{n}$) to the IM boundary, which is given by {\color{black}the} gradient to the surface at that point. The angle of incidence is then given by $i=\pi-\cos^{-1}(\hat{n}\cdot\hat{k})$ (Figure \ref{fig:ini_k_IM}).

%To find the angle of refraction, we need the refractive indices both outside and inside the inner magnetosphere (IM). 
We now determine the angle of refraction. As already mentioned above, the refractive index outside the IM ($\mu_1$) is assumed to be unity and that inside the IM ($\mu_2$) can be calculated using the density model of the IM. One caveat here is that $\mu_2$ depends not only on the plasma density, but also on the angle between the wave vector and the magnetic field vector immediately inside the IM (Eq. \ref{eq:refractive_index}). It is not possible to know the direction of the wave vector inside the IM beforehand. {\color{black} This difficulty can be overcome by adopting an iterative approach. We, first calculate the value of $\mu_2$ using the same value for the angle between the wave vector and the magnetic field as that before entering the IM.}
%It is not possible to know the direction of the wave vector inside the IM beforehand, we, therefore, use the same value for the angle between the wave vector and the magnetic field as that before entering the IM. This strategy is justified because the change of $\mu$ between immediately outside and inside the IM happens predominantly due to change in number density. 
Using the angle of incidence ($i$), $\mu_1$ and $\mu_2$, the angle of refraction $\theta_r$ can be found from {\color{black}the} Snell's law. To find the direction of the wave vector just after entering the IM ($\textbf{k}_\mathrm{IM}$), we note that:
\begin{align}
\textbf{k}_\mathrm{IM}\propto \hat{r}+r\theta^\prime\hat{\theta}+r\sin\theta\phi^\prime\hat{\phi}\label{eq:k_vect_inside_IM}
\end{align}

For obtaining the ray path, we are interested only in the direction of $\textbf{k}_\mathrm{IM}$. Hence we set {\color{black}$\textbf{k}_\mathrm{IM}=\gamma(\hat{r}+r\theta^\prime\hat{\theta}+r\sin\theta\phi^\prime\hat{\phi})$, where $\gamma=\pm 1$}. We already know the value of $r$, $\theta$ and $\phi$ at point $Q$, which are respectively $r_0$, $\theta_0$ and $\phi_0$. To find $\theta^\prime=d\theta/dr$ and $\phi^\prime=d\phi/dr$ at point ($r_0$, $\theta_0$, $\phi_0$) inside the IM, we solve the following two equations (Figure \ref{fig:ini_k_IM}):
\begin{align*}
\hat{k}\cdot\textbf{k}_\mathrm{IM}&=|\textbf{k}_\mathrm{IM}|\cos(\theta_r-i)\\
\hat{n}\cdot\textbf{k}_\mathrm{IM}&=|\textbf{k}_\mathrm{IM}|\cos(\pi-\theta_r)
\end{align*}
{\color{black}By solving these two equations, we can find $\theta^\prime_0$ and $\phi^\prime_0$ immediately after entering the IM. Once the new direction of the wave vector is known, we can find the angle made by it with the local magnetic field vector. This in turn is used to calculate $\mu_2$ to get an improvement over the previous estimation. This cycle is continued until we achieve convergence. $\theta^\prime_0$ and $\phi^\prime_0$ are then calculated using the value of $\mu_2$ obtained after convergence.}

The five values: $r_0$, $\theta_0$, $\phi_0$, $\theta^\prime_0$ and $\phi^\prime_0$ will serve as {\color{black} the} initial conditions which will be required to solve the ray path inside the IM.

\subsection{Ray path inside the inner magnetosphere}\label{subsec:ray_path_inside_IM}
To find the ray path inside the IM, we  use {\color{black} the} Fermat's principle \citep{wolin1953}. According to this principle, the path is given by the  minimization of $\int\mu ds$  (note that $\mu=\mu_2$ at the point $Q$ inside the  IM). Now,
\begin{align*}
ds&=\sqrt{dr^2+r^2d\theta^2+r^2\sin^2\theta d\phi^2}\\
&=dr\sqrt{1+r^2\theta^{\prime 2}+r^2\sin^2\theta\phi^{\prime 2}}\\
&=G\times dr
\end{align*}
Where, $G=\sqrt{1+r^2\theta^{\prime 2}+r^2\sin^2\theta\phi^{\prime 2}}$. Let, 
%Since we assume the number density to be dependent only on $r$, plasma frequency $\omega_p$ is a function of $r$ only. 

\begin{align*}
F(r,\theta,\phi,\theta^\prime,\phi^\prime)=\mu\times G
\end{align*}
To minimize $F$, we have from {\color{black} the} Euler-Lagrange's equation:

\begin{align}
&\frac{\partial F}{\partial \theta}-\frac{d}{dr}\frac{\partial F}{\partial \theta^\prime}=0\label{eq:ray_path1}\\
&\frac{\partial F}{\partial \phi}-\frac{d}{dr}\frac{\partial F}{\partial \phi^\prime}=0\label{eq:ray_path2}
\end{align}
By integrating this equation numerically, we can find the ray path \{$\theta(r),\phi(r)$\} inside the IM. To do that, we cast the above equations in the following form:
\begin{align*}
\frac{dY_1}{dr}&=\frac{\partial F}{\partial \theta} , \quad \frac{dY_2}{dr}=\frac{\partial F}{\partial \phi}  \\
Y_1&=\frac{\partial F}{\partial \theta^\prime} , \quad Y_2=\frac{\partial F}{\partial \phi^\prime}
\end{align*}
Assuming a step size of the integration (in $r$) to be $\Delta r$. Then at a step $i$, we have:
\begin{align*}
r_{i+1}&=r_i+\Delta r\\
\theta_{i+1}&=\theta_i+\theta^\prime_i\Delta r\\
\phi_{i+1}&=\phi_i+\phi^\prime_i\Delta r\\
Y_{1 ,i+1}&=Y_{1,i}+\frac{dY_1}{dr}\biggr |_{i}\Delta r\\
Y_{2 ,i+1}&=Y_{2,i}+\frac{dY_2}{dr}\biggr |_{i}\Delta r
\end{align*} 
To obtain $\theta^\prime_{i+1}$ and $\phi^\prime_{i+1}$, we minimize the following equations:
\begin{align*}
&Y_{1,i+1}-\left(\frac{\partial F}{\partial \theta^\prime}\right)_{i+1}=0,\quad Y_{2,i+1}-\left(\frac{\partial F}{\partial \phi^\prime}\right)_{i+1}=0
\end{align*}
Once we find $\theta^\prime_{i+1}$ and $\phi^\prime_{i+1}$, we can obtain $(dY_1/dr)_{i+1}$ and $(dY_2/dr)_{i+1}$ in the following way:
\begin{align*}
&\frac{dY_1}{dr}\biggr |_{i+1}=\left(\frac{\partial F}{\partial \theta}\right)_{i+1}, \quad \frac{dY_2}{dr}\biggr |_{i+1}=\left(\frac{\partial F}{\partial \phi}\right)_{i+1}
\end{align*}

We stop the integration when we achieve the condition: $r/\sin^2\theta\geq R_\mathrm{A}$. 
%The details of the integration are given in the appendix.

\subsection{Ray direction after exiting the IM}\label{subsec:ray_out}
\begin{figure}
\centering
\includegraphics[trim={10cm 10cm 0cm 0cm}, clip, width=0.45\textwidth]{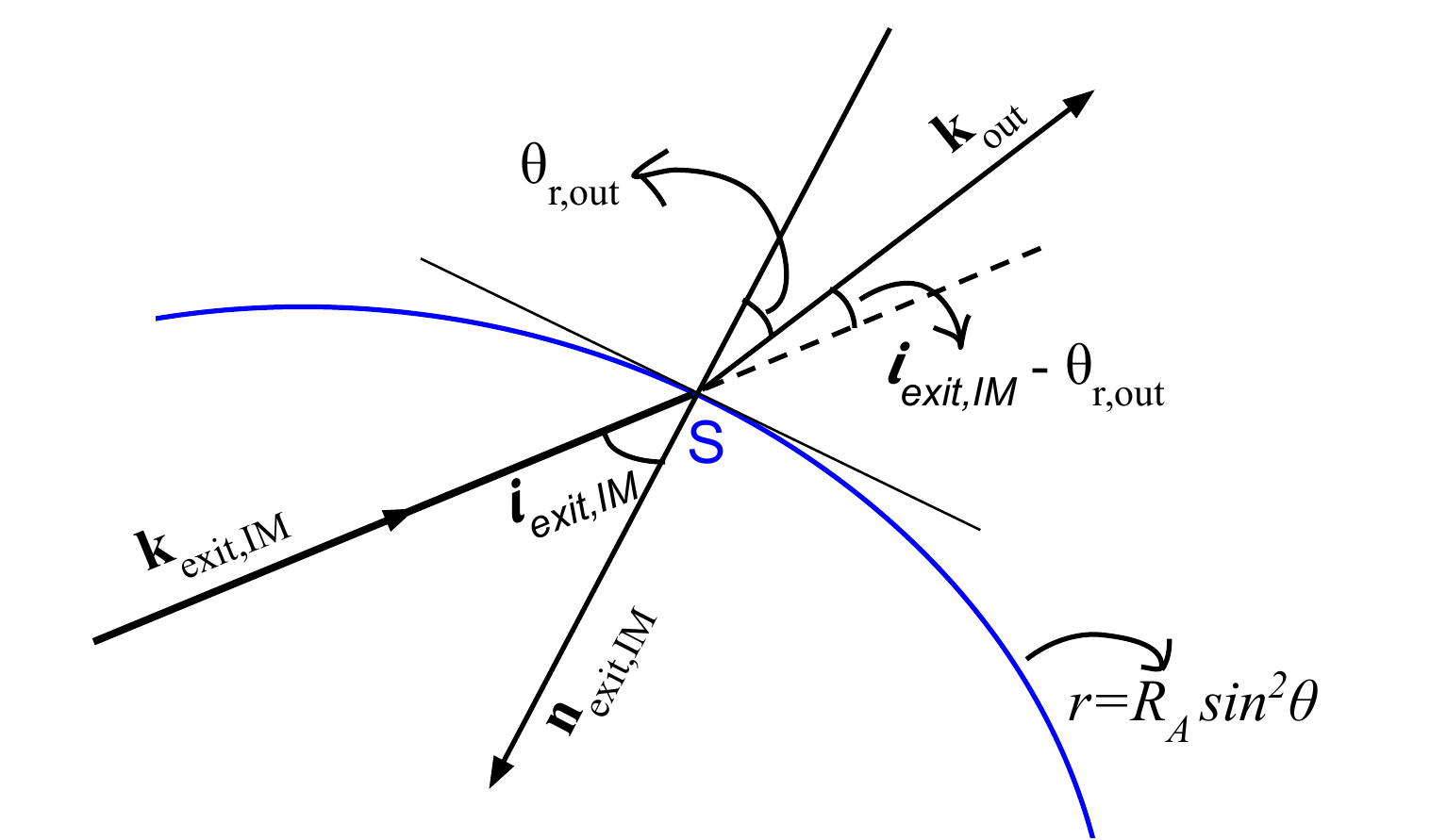}
\caption{The relation between the direction of the wave vector after entering the IM ($\textbf{k}_\mathrm{out}$), the direction of the wave vector just insider the IM
 ($\textbf{k}_\mathrm{exit,IM}$) and the direction of the normal to the boundary of the inner magenetosphere ($\textbf{n}_\mathrm{exit,IM}$). \label{fig:IM_exit}}
\end{figure}

Let the point at which the ray exits the IM be $S(r_\mathrm{exit,IM},\theta_\mathrm{exit,IM},\phi_\mathrm{exit,IM})$ (Figure \ref{fig:IM_exit}). The value of $\theta^\prime$ and $\phi^\prime$ inside {\color{black} the} IM at point $S$ be $\theta^\prime_\mathrm{exit,IM}$ and $\phi^\prime_\mathrm{exit,IM}$ respectively. Using these values, we can readily obtain the direction of the wave vector inside the IM at the exit point $\hat{k}_\mathrm{exit,IM}$ (Eq. \ref{eq:k_vect_inside_IM}). To find out the wave vector direction after exiting the IM $\hat{k}_\mathrm{out}$, we need the angle of refraction. This can be found out using a procedure similar to the one adopted to find the angle of refraction at the time of entering the IM (described in \S\ref{subsec:ini_cond}). $\hat{k}_\mathrm{out}$ can then be found by solving the following two equations:
\begin{align*}
\hat{k}_\mathrm{exit,IM}\cdot\hat{k}_\mathrm{out}&=\cos(i_\mathrm{exit,IM}-\theta_\mathrm{r,out})\\
\hat{n}_\mathrm{exit,IM}\cdot\hat{k}_\mathrm{out}&=\cos(\pi-\theta_\mathrm{r,out})
\end{align*}
where $i_\mathrm{exit,IM}$ and $\theta_\mathrm{r,out}$ are respectively the angle of incidence and angle of refraction at point $S$, $\hat{n}_\mathrm{exit,IM}$ is the inward normal to the IM at point $S$.

\section{Application of the model}\label{sec:output}
In this section, we demonstrate applicability of this framework in studying various properties of the ECME. We consider two kinds of density distributions in the IM: the first is the case of an azimuthally symmetric density distribution in the IM, and the second is the case of an azimuthally asymmetric density distribution in the IM. For the simulation presented here, the step-size was determined in an ad-hoc manner, which is that we varied the step-size of integration until we get no significant change in the results with further decrease in the step-size. For practical purpose, we recommend to use adaptive step-size, which can be obtained by determining the length scale of {\color{black} the} change in the number density at a given point in the IM.

\subsection{ECME from a star with an azimuthally symmetric magnetosphere}\label{subsec:azimuth_sym}
\begin{figure}
\centering
\includegraphics[width=0.5\textwidth]{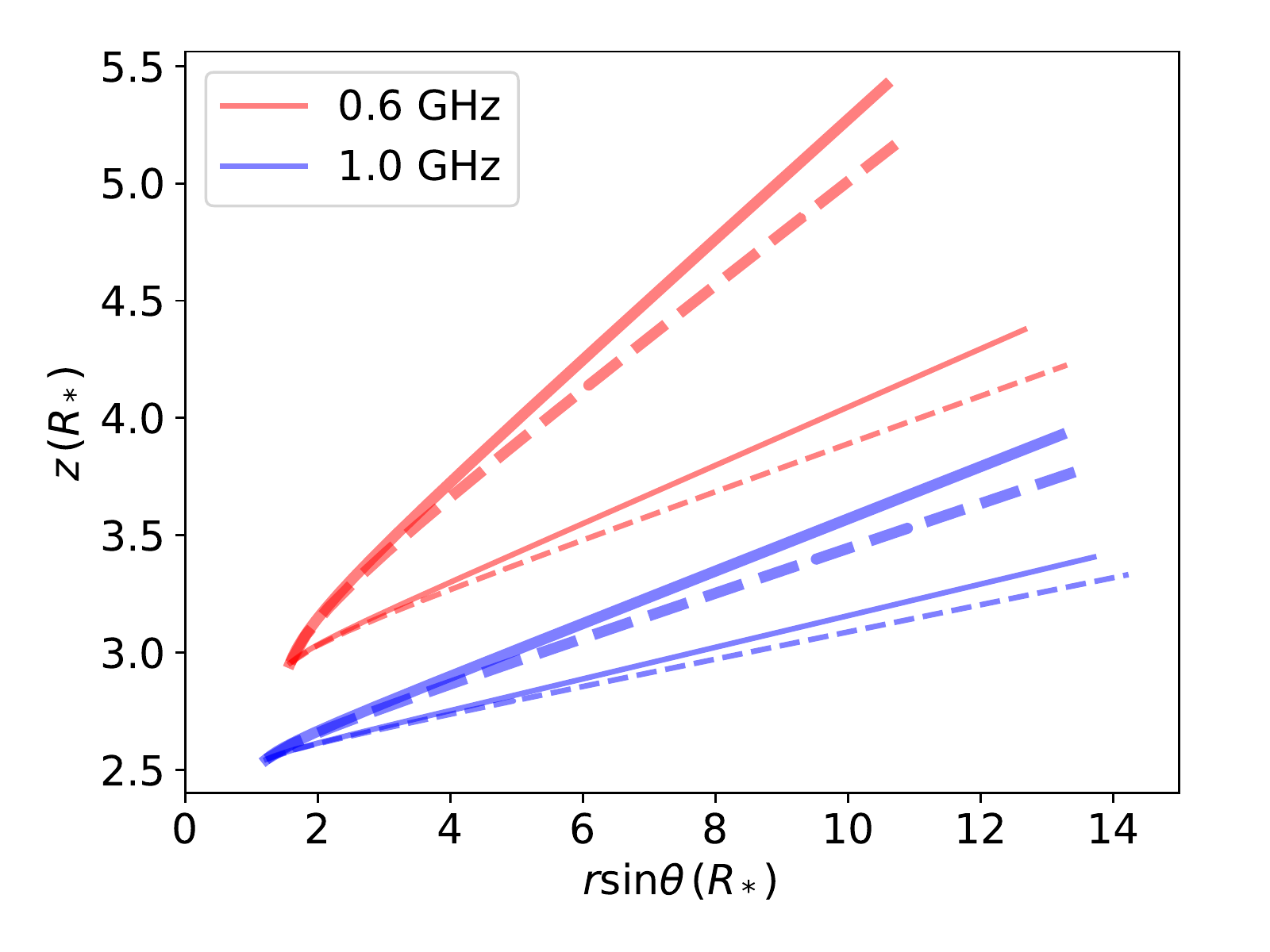}
\caption{Ray paths inside the inner magnetosphere (IM) generated by our code (solid lines) and those obtained by assuming single refraction at the boundary between inner and middle magnetosphere (dashed lines). The thick and thin lines are respectively for extra-ordinary (X-) and ordinary (O-) mode of emission at the second harmonic. We have assumed a radially decreasing density profile inside IM given by $n_p=n_\mathrm{p0}/r$, where $n_\mathrm{p0}=10^9$ cm$^{-3}$. The values of the other parameters used here are $B_0=4$ kG, $R_\mathrm{A}=15.0$ and $L=18.0$. All the distances are in the units of stellar radius.  \label{fig:ray_path_inside_IM}}
\end{figure}

\begin{figure}
\centering
\includegraphics[width=0.45\textwidth]{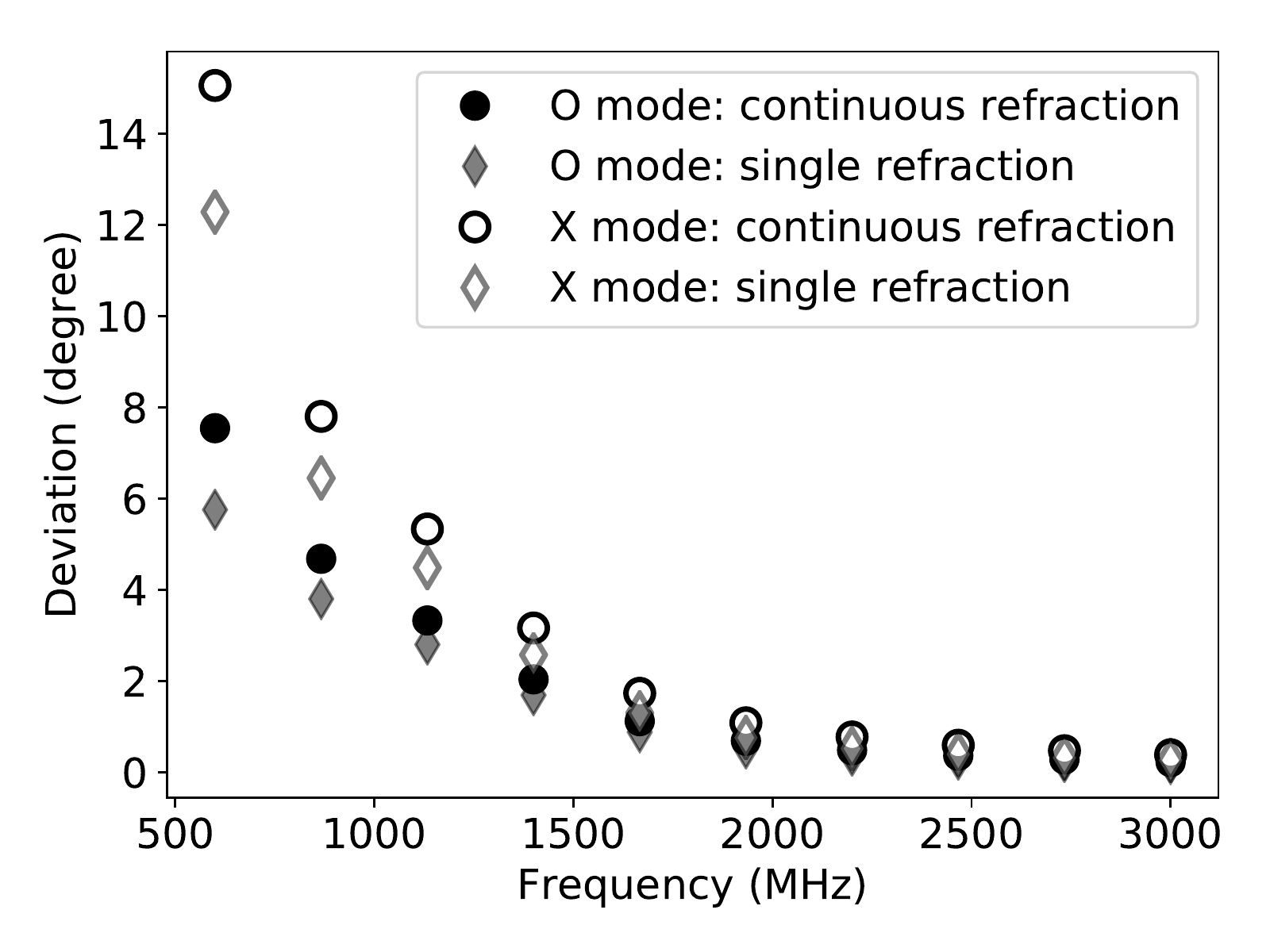}
\caption{Deviation of the ECME after exiting the IM w.r.t. the initial direction of emission vs frequency of the radiation. The circles correspond to the ones obtained from our code, and the diamonds correspond to the deviations obtained by considering single refraction only. The values of other parameters are same as that in Figure \ref{fig:ray_path_inside_IM}. \label{fig:deviation_vs_freq}}
\end{figure}

We consider a star with an axi-symmetric dipolar magnetic field. The star has an inclination angle of 46.5$^\circ$ and obliquity of 76$^\circ$ (close to the values for CU\,Vir). We assume that the density inside the IM is $n_p=n_{p0}/r$, where $r$ is the distance in the units of stellar radius from the centre of the star and $n_{p0}$ is a normalization constant with $n\to n_{p0}$ as $r\to 1$ \citep[expected from stellar rotation,][]{leto2006}. We also assume that the ECME is produced at the second harmonic, i.e. $s=2$. For this type of a distribution, {\color{black} the} ECME emitted at each point of a given auroral circle experience identical densities and follow identical paths, so that none of the observable properties of ECME has any dependence on the magnetic azimuthal coordinate $\phi$ (will be called $\phi_\mathrm{B}$ from now onwards). 

%We first check whether the ray paths calculated by our code are any different from the ones obtained by assuming single refraction suffered by ECME at the time of entering the IM. For that, 
To calculate the ray paths for the ECME, we assume $n_{p0}=10^9\,\,\mathrm{cm^{-3}}$ and $R_\mathrm{A}=15\,R_*$, where $R_*$ is the stellar radius.
%, and consider two frequencies : 600 MHz and 1 GHz. We also assume that 
We also assume that the ECME is produced in the auroral rings formed of magnetic field lines with equatorial radius $L=18\,R_*$. We show the results of our simulation at two frequencies, 0.6\,GHz and 1\,GHz,  in Figure \ref{fig:ray_path_inside_IM}. We compare our results  with the ones obtained by considering a single refraction at the boundary between {\color{black} the} middle and {\color{black} the} inner magnetosphere (dashed lines), as per the framework of \citet{lo2012}. 
We carry out simulations for both extra-ordinary (X-) and ordinary (O-) modes.
We find that although the single refraction approach can produce the qualitative picture quite nicely, it underestimates the deviation suffered by the ray. 
%We find that in either case, the single refraction treatment underestimates the deviation suffered by the ray (Figure \ref{fig:ray_path_inside_IM} ). in other words, to produce the same amount of net deviation of the ECME, the single refraction model will require a higher value of $n_{p0}$. However the qualitative differences between ray paths at the two frequencies, or at the two magneto-ionic modes, are the same for both approaches (i.e. the single refraction approach, and our continuous refraction approach). 
For example, both indicate that the lower frequency deviates larger than the higher frequency at a given magneto-ionic mode; and at a given frequency, radiation at the X-mode deviates more than that at the O-mode. This is more evident from Figure \ref{fig:deviation_vs_freq} where we plot the deviation angle (defined as $\theta_D=\pi/2-\cos^{-1}(\hat{k}_\mathrm{out}\cdot\hat{z})$, i.e. the angle made by the ray direction after exiting the IM with the magnetic equatorial plane) vs {\color{black} the} frequency of the radiation. We see that as the frequency of {\color{black} the} radiation increases, the deviation suffered decreases, and also the difference in {\color{black} the} deviation obtained from our code and that obtained by assuming single refraction at the time of entering the IM decreases. This is expected because as we go to higher frequencies (i.e. going closer to the star), the ratio $\nu_\mathrm{p}/\nu$ ($\nu_\mathrm{p}$ is the plasma frequency) becomes smaller. Note that this phenomenon of $\nu_\mathrm{p}/\nu$ decreasing with decreasing $r$ is a consequence of the assumed density profile, for which $\nu_\mathrm{p}\propto 1/\sqrt{r}$, whereas $\nu\propto \nu_\mathrm{B}\propto 1/r^3$ ($\nu_\mathrm{B}$ is the electron gyrofrequency at the site of emission) so that $\nu_\mathrm{p}/\nu\propto r^{2.5}$.

From the point of view of an observer, the ray paths are not directly measurable. We need a quantity which is directly measurable from observations and then can be compared with the values predicted by a model. One such quantity is the difference in rotational phases of arrival of pulse at different frequencies  (we refer to this quantity as `lag').
For our simple density profile, after obtaining  $\hat{k}_\mathrm{out}$ for any $\phi_{01}$ (\S\ref{subsec:ray_out}), we can obtain the deviation angle, which we define as $\theta_D=\pi/2-\cos^{-1}(\hat{k}_\mathrm{out}\cdot\hat{z})$. Here $\pi/2$ is the angle between the dipole axis and the original direction of emission and $\cos^{-1}(\hat{k}_\mathrm{out}\cdot\hat{z})$ is the angle between the dipole axis and the direction of the radiation after exiting the IM. This angle is related to the rotational phase $\phi_\mathrm{rot}$ with the following relation \citep{trigilio2000}:

\begin{align}
\sin\theta_D=&\sin\beta\sin \alpha\cos2\pi(\phi_\mathrm{rot}-\phi_\mathrm{rot,0})+\nonumber\\& \cos\beta\cos\alpha\nonumber \\
\Rightarrow \phi_\mathrm{rot}-\phi_\mathrm{rot,0}=&\frac{1}{2\pi}\cos^{-1}\left(\frac{\sin\theta_D-\cos\beta\cos\alpha}{\sin\beta\sin\alpha}\right)\label{eq:1}
\end{align}

where $\alpha$ and $\beta$ are the inclination angle and the angle between the rotation axis and magnetic axis, respectively, $\phi_\mathrm{rot,0}$ is the reference rotational phase (which corresponds to the rotational phase when the line of sight component of the magnetic field is maximum). Let the two frequencies be $\nu_1$ and $\nu_2$. The rotational phases of arrival are $\phi_\mathrm{rot,1}$ and $\phi_\mathrm{rot,2}$ respectively and $\phi_\mathrm{rot,2}-\phi_\mathrm{rot,1}=\Delta \phi_\mathrm{rot}$. We have from Eq.\ref{eq:1}:
\begin{align}
\Delta\phi_\mathrm{rot}&=(\phi_\mathrm{rot,2}-\phi_\mathrm{rot,0})-(\phi_\mathrm{rot,1}-\phi_\mathrm{rot,0})\nonumber\\
&=\frac{1}{2\pi}\cos^{-1}\left(\frac{\sin\theta_{D2}-\cos\beta\cos\alpha}{\sin\beta\sin\alpha}\right)\nonumber\\
&\phantom{{}=1}-\frac{1}{2\pi}\cos^{-1}\left(\frac{\sin\theta_{D1}-\cos\beta\cos\alpha}{\sin\beta\sin\alpha}\right) \label{eq:3_0}
\end{align}

Thus from the calculated $\theta_D$s for a pair of frequencies, the corresponding lag (=$\Delta\phi_\mathrm{rot}$) can be obtained from Eq. \ref{eq:3_0} which can be compared to observations.

\begin{figure*}
\centering
\includegraphics[trim={0cm 5cm 0cm 0cm}, clip, width=0.85\textwidth]{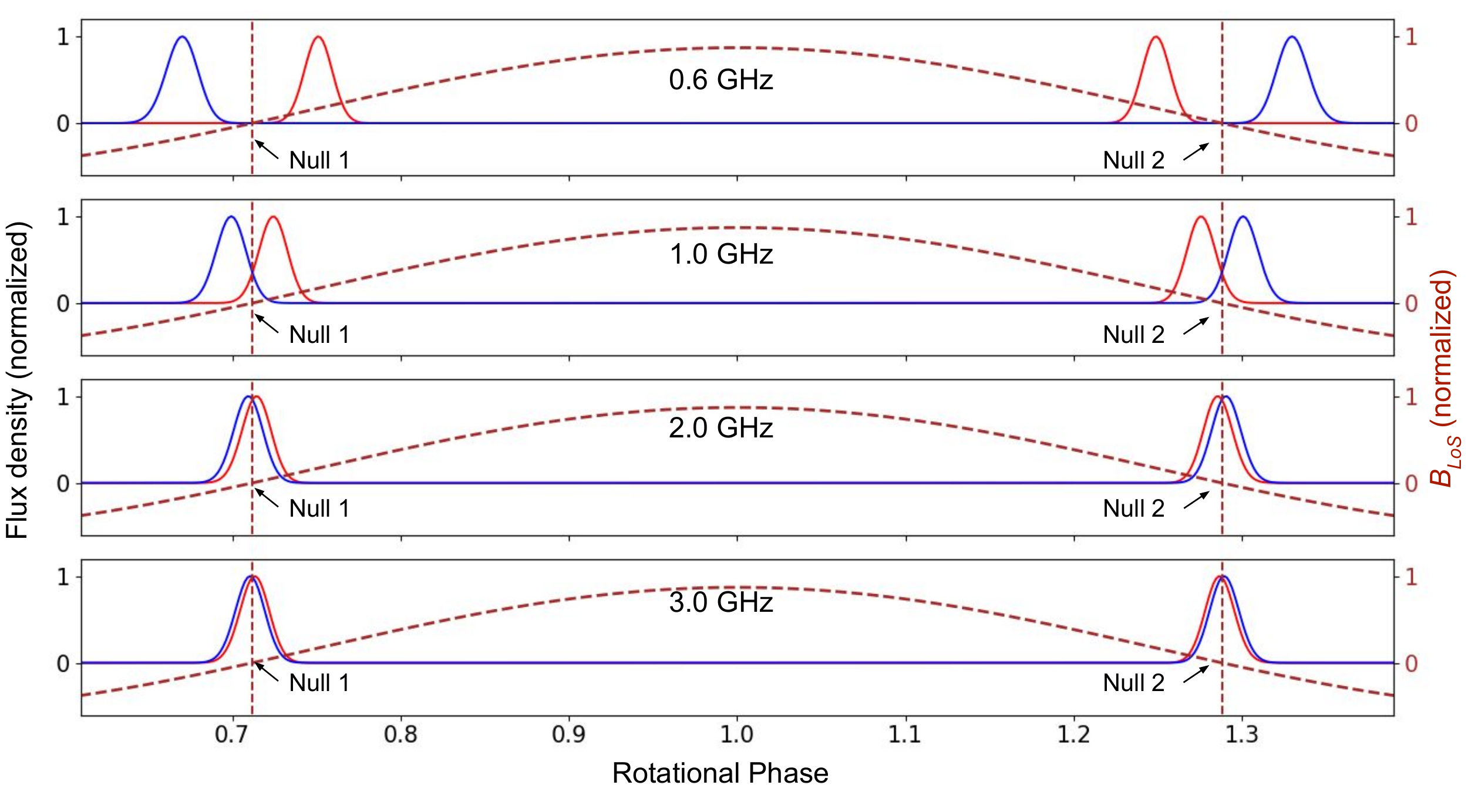}
\caption{The normalized lightcurves for ECME at 1 GHz, 2 GHz and 3 GHz, emitted at the second harmonic in the extra-ordinary mode. The red and blue correspond to ECME produced at the north and south magnetic hemispheres respectively. The density profile in the IM is  given by $n_p=n_{p0}/r$, where $n_{p0}=10^9$ $\mathrm{cm^{-1}}$. We have used $\sigma_\theta=3^\circ$ (\S\ref{sec:ecme_lightcurve}). We also show the normalized line of sight component of the stellar magnetic field ($B_\mathrm{LoS}$, brown dashed curve). We have used a polar magnetic field strength of 4 kG, inclination angle of 46.5$^\circ$, obliquity of $76^\circ$.\label{fig:lightcurves_sym_IM}}
\end{figure*}

We end this subsection by showing the X-mode ECME lightcurves at four different frequencies, emitted at the second harmonic (Figure \ref{fig:lightcurves_sym_IM}). The details of how we obtain the lightcurves from our code is given in the appendix. We have assumed that $B_0=4$ kG, $\alpha=46.5^\circ$, $\beta=76^\circ$ and $n_{p0}=10^9\,\,\mathrm{cm^{-3}}$. We have also assumed a Gaussian profile for the ECME pulses (\S\ref{sec:ecme_lightcurve}) with $\sigma_\theta=3^\circ$ (this corresponds to a full width at half maxima of $\approx 7^\circ$). The values of the flux densities at a given frequency are normalized by the maximum value at that frequency. We also show normalized \bz\, in Figure \ref{fig:lightcurves_sym_IM}. For convenience, we name the magnetic null, at which \bz~changes from negative to positive, as null 1 and the other magnetic null, where \bz~changes from positive to negative as null 2.
We find that near null 1, the pulse from the south magnetic hemisphere (blue curve) arrives ahead of those from the north magnetic hemisphere (red curve), and vice-versa. Note that this characteristic was also obtained by \citet{leto2016} in which (effectively) the single refraction scenario was used. We also find that the separation between pulses around a given null decreases as we go to higher frequencies. This frequency dependent separation is a strong function of density in the IM. The proposed framework will be useful if one wants to exploit this dependence to estimate density in the IM.

%We find that the resulting lightcurves are qualitatively similar to the ones simulated by \citet{leto2016} in which (effectively) the single refraction scenario was used. We also find that near null 1, the pulse from the south magnetic hemisphere (blue curve) arrives ahead of those from the north magnetic hemisphere (red curve), and vice-versa, consistent with the single refraction model. The separation between pulses around a given null decreases as we go to higher frequencies. However for quantitative treatment, e.g., if one wants to use multifrequency observations of ECME to estimate density in the IM, the framework presented here will be more suitable.

\subsection{ECME from a star with an azimuthally asymmetric magnetosphere}\label{subsec:azimuth_asym}
Stars with high obliquity are predicted to have overdense regions in its magnetosphere, which are not azimuthally symmetric \citep[the `Rigidly Rotating Model', or `RRM',][]{townsend2005}. The deviation from the azimuthal symmetry is most severe for $\beta$ close to $90^\circ$. Among the stars known to produce ECME, many have high obliquities, e.g. {\color{black} the} magnetic B star HD\,142990 has $\beta\approx 84^\circ$ \citep{shultz2019_3}. In this section, we carry out our 
simulation to ascertain the effect of this azimuthally asymmetric density distribution on the ECME properties of the star over a  wide range of frequencies. 
%In addition to that, it is also important to have an idea about what kind of changes, ECME at a given frequency will undergo, for a particular type of density distribution. In the subsequent part of this subsection, we address the latter concern.

%we consider a azimuthally asymmetric density distribution inside IM, and show how the lightcurves at different frequencies get affected by that.

\begin{figure*}
\centering
\includegraphics[width=0.49\textwidth]{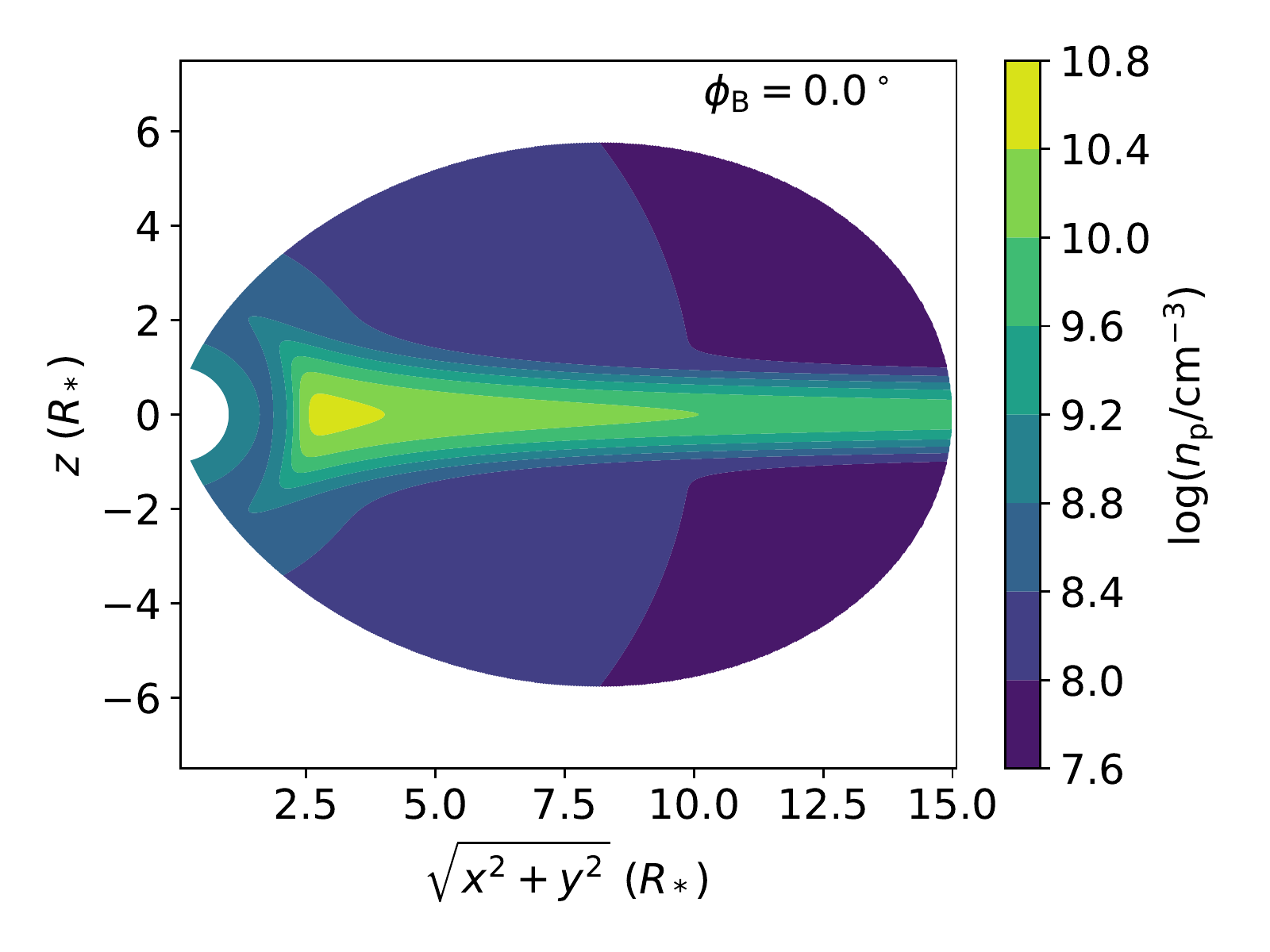}
\includegraphics[width=0.49\textwidth]{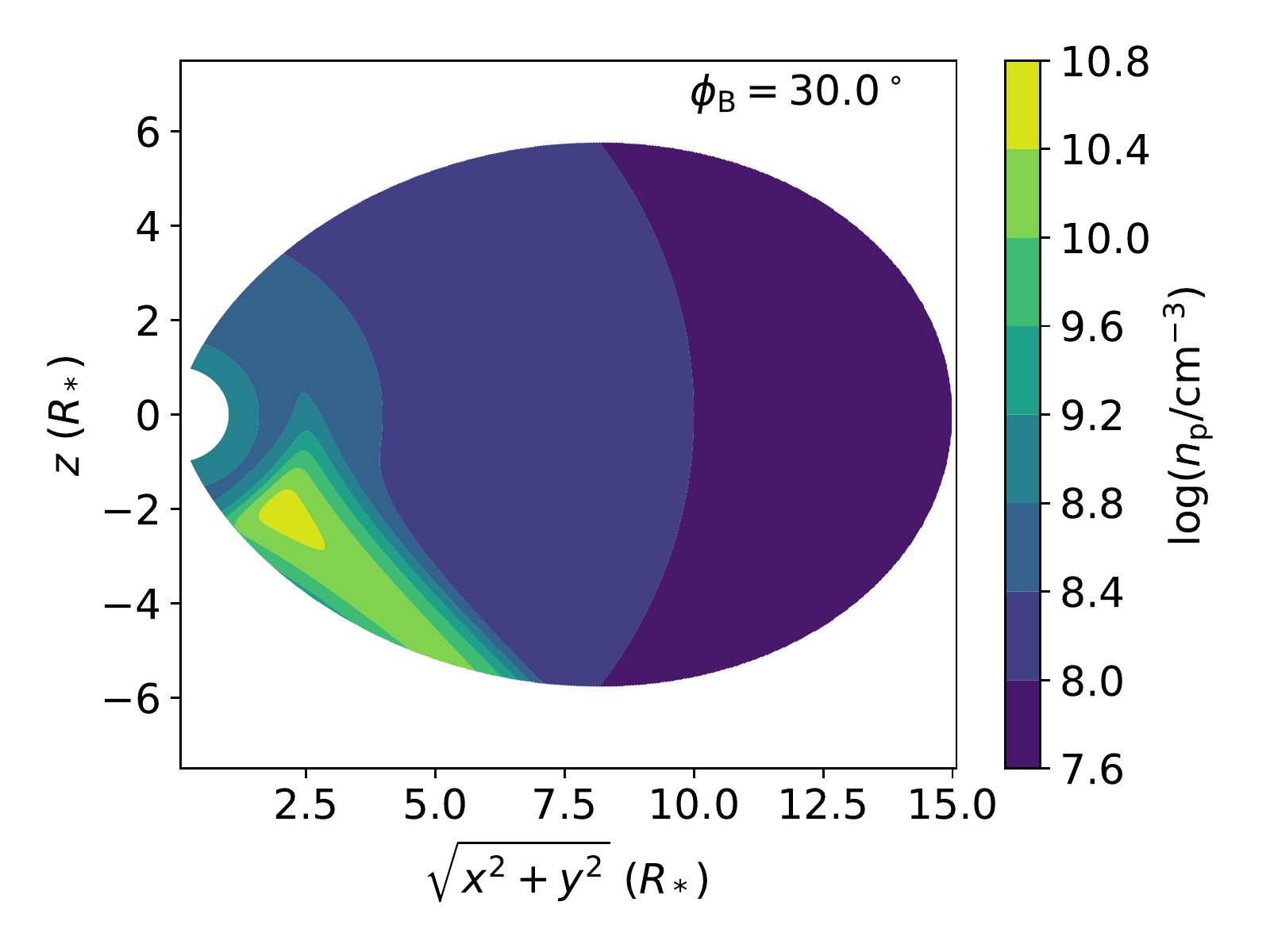}
\caption{The density profile given by Eq. \ref{eq:complex_asym_IM} for two different values of the magnetic azimuthal coordinate $\phi_\mathrm{B}$.\label{fig:complex_IM_asym}}
\end{figure*}

The density profile that we will take as an example is inspired from the RRM model of \citet{townsend2005}, and is given by the following equation:

\begin{align}
n_p&=\frac{n_{p0}}{r}\left(1+100\exp\left(-\frac{3\tilde{z}^2}{\sigma(\tilde{x})^2}\right)(1-\Delta)\right)\label{eq:complex_asym_IM}\\
\tilde{z}&=r\sin(\theta-\theta_0),\quad \tilde{x}=r\cos(\theta-\theta_0)\nonumber\\
\theta_0&=\frac{\pi}{2}(1+\sin\phi_\mathrm{B}),\quad\sigma(\tilde{x})=0.7\exp\left(\frac{R_\mathrm{A}}{R_\mathrm{A}+\tilde{x}^2}\right)\nonumber\\
\Delta(r)&=\frac{1}{1+\exp\{2M(r-r_0)\}},\quad r_0=2.5\nonumber
\end{align}

Note that the equations are written for the magnetic frame of reference. This density profile is shown in Figure \ref{fig:complex_IM_asym} for two values of $\phi_\mathrm{B}$: $0^\circ$ and $30^\circ$.

\begin{figure*}
\centering
\includegraphics[trim={0cm 4.5cm 0cm 0cm}, clip, width=0.85\textwidth]{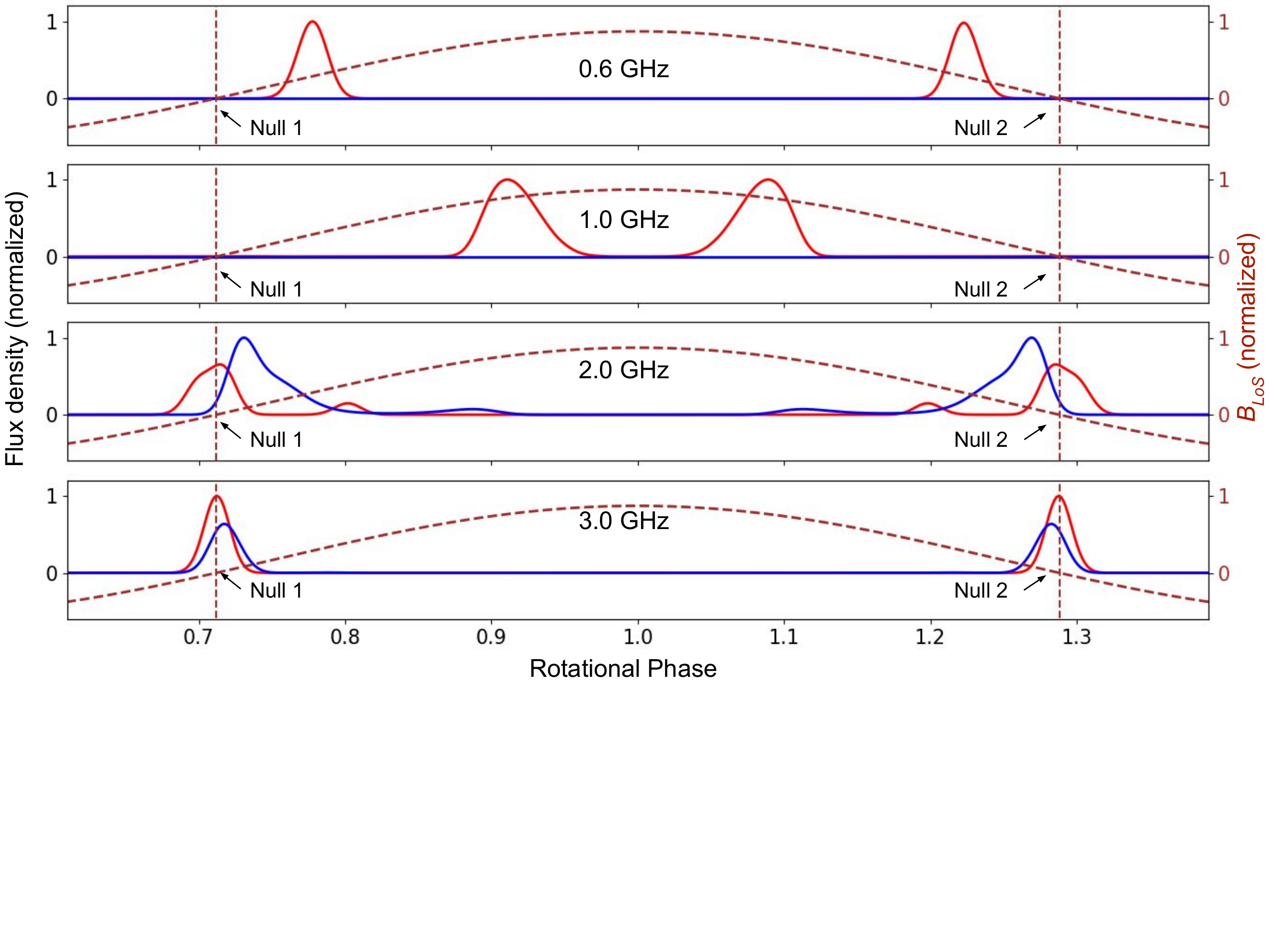}
\caption{Normalized lightcurves obtained for the IM density profile given by Eq. \ref{eq:complex_asym_IM}, with $\sigma_\theta=3^\circ$. The red and blue correspond to ECME produced at the north and south magnetic hemispheres respectively. The brown dashed curves correspond to the normalized line of sight magnetic field.\label{fig:lightcurve_asym_IM}. }
\end{figure*}

%\authorcomment1{I would prefer to be consistent in this section also and add 0.6 GHz here similar to section 4.1. I also think it would be informative to add Fig 6 equivalent for this asymmertic magnetosphere with complex density profile.}
We again consider ECME at 0.6, 1, 2  and 3 GHz, emitted at {\color{black} the} X-mode, in the second harmonic. In Figure \ref{fig:lightcurve_asym_IM}, we show the normalized output lightcurves along with \bz~(brown dashed curve) for $\sigma_\theta=3^\circ$. These lightcurves are different from those in Figure \ref{fig:lightcurves_sym_IM}, or, those simulated by \citet{leto2016} in many important aspects, such as:
%These lightcurves are certainly not the ideal ones expected from a star with an azimuthally symmetric magnetosphere. These unusual properties are:
\begin{enumerate}
\item At 0.6 and 1 GHz, there is no pulse from the south magnetic hemisphere.
\item Between 0.6 and 1 GHz, the offset of the pulse from the magnetic null is higher at the higher frequency, unlike the case for the azimuthally symmetric IM (\S\ref{subsec:azimuth_sym}, Figure \ref{fig:lightcurves_sym_IM}).
\item At 2 and 3 GHz, the strengths of the pulses from the north and {\color{black} the} south magnetic hemispheres are different despite taking perfectly dipolar magnetic field.
\item At 2 GHz, there are weaker secondary pulses, in addition to the primary pulse.
\item The sequence of arrival of pulses at 2 and 3 GHz from the opposite magnetic hemispheres are opposite to what one expects for the azimuthally symmetric case (Figure \ref{fig:lightcurves_sym_IM}).
\item The magnetic null phases do not lie at the mid-point between {\color{black} the} pulses from the opposite magnetic hemispheres.
\end{enumerate}

We first address the last point, which is straightforward to understand. Due to the asymmetry of the  magnetosphere, 
 %which is that the magnetic nulls do not lie at the midpoint between pulses from opposite magnetic hemisphere (for the cases where there are observable pulses from both hemispheres). This is because now 
 the radiation produced at the north and {\color{black} the} south magnetic hemispheres will undergo different amount of deviation, resulting in a loss of symmetry about the magnetic null.

The reason for the absence of ECME pulse from south magnetic hemisphere at 0.6 and 1 GHz is that for the density distribution considered here, many of the rays produced at the relevant auroral circles, cannot pass through the IM because of the very high density for which the refractive index becomes imaginary. This happens for the rays originated at the south magnetic hemispheres.
% that would have contributed to the observable pulses. 
Note that an order of magnitude estimate of the critical density, above which radiation at 0.6 and 1 GHz cannot pass is  $5\times10^{9}\,\,\mathrm{cm^{-3}}$ and $10^{10}\,\,\mathrm{cm^{-3}}$ respectively which is achievable for the density profile given by Eq. \ref{eq:complex_asym_IM}. 

The higher offset of ECME pulses from the magnetic null at higher frequencies between 0.6 and 1 GHz is due to the fact that the density experienced by ECME at 1 GHz is much higher than that for 0.6 GHz.

The third point about different strengths for pulses from north and south magnetic hemispheres is actually not surprising.  In our framework
 the strength of an ECME pulse is determined by three factors: 1) the number of rays that contribute to the pulse, 2) the difference in deviations suffered by the contributing rays and 3) the angles made by the contributing rays with the LoS at their closest proximities. Larger the number of {\color{black} the} contributing rays, and smaller the angle between the contributing rays and the LoS at their closest encounters, the higher will be the corresponding pulse strength. On the other hand, larger the difference in deviations suffered, lower will be the pulse height and broader will be the pulse, since the contributing rays reach the observer over a relatively large range of rotational phases. For the density distribution that we have considered, these three factors need not be same for pulses from opposite magnetic hemispheres, or at different frequencies, resulting in the discrepancies mentioned above.
%not much useful from the perspective of an observer in terms of extracting information about the density in the magnetosphere. This is because, we often see that 
However, an additional factor causing  ECME of different strengths at different magnetic hemispheres is the instability at the emission site \citep[e.g.][ Das et al. in prep]{trigilio2011,das2018}. Thus the different strengths of ECME pulses do not give much insight in terms of extracting information about the density in the magnetosphere. 
%both pulse strength and pulse shapes of ECME change over a short timescale \citep[e.g.][ Das et al. in prep]{trigilio2011,das2018} which indicates that these two observed properties are dominated by instability at the emission site. 

\begin{figure*}
\centering
\includegraphics[width=0.85\textwidth]{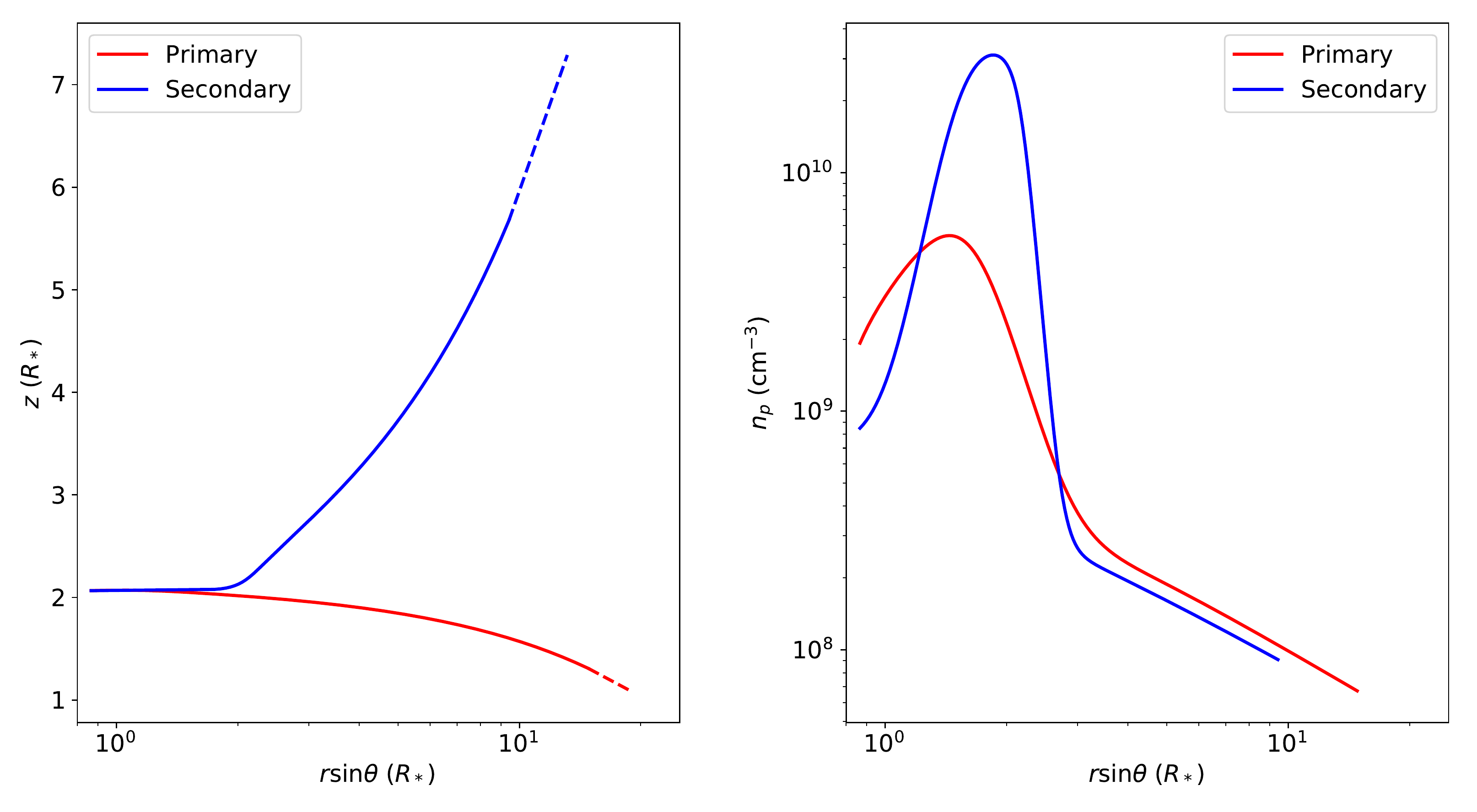}
\caption{\textit{Left:} The ray paths that contribute to the 2 GHz pulse (X-mode) from the north magnetic hemisphere for azimuthally asymmetric density distribution. The red and blue correspond to the primary and secondary components of the pulse (Figure \ref{fig:lightcurve_asym_IM}, for details, see \S\ref{subsec:azimuth_asym}). \textit{Right:} The corresponding number densities along the paths inside IM.\label{fig:ray_path_ne_asym_IM_2GHz}}
\end{figure*}

%Although the difference in pulse strengths and shapes produced by propagation effect may not be of much interest to an observer, 
The existence of weaker secondaries at a particular frequency (here at 2 GHz) too is a consequence of the fact that the three factors (described in the preceding paragraph) assume different values for different frequencies and different magnetic hemispheres. To illustrate the difference in ray paths, we show one of the contributing rays to the primary and one of those to the secondary pulse from the north magnetic hemisphere at 2 GHz in the left panel of Figure \ref{fig:ray_path_ne_asym_IM_2GHz}. We also show the corresponding densities encountered along the ray paths in the right panel. We further find that the number of contributing rays to the secondary is smaller than that for the primary, which is expected from the observation that the latter is much stronger than the former.

\begin{figure*}
\centering
\includegraphics[width=0.88\textwidth]{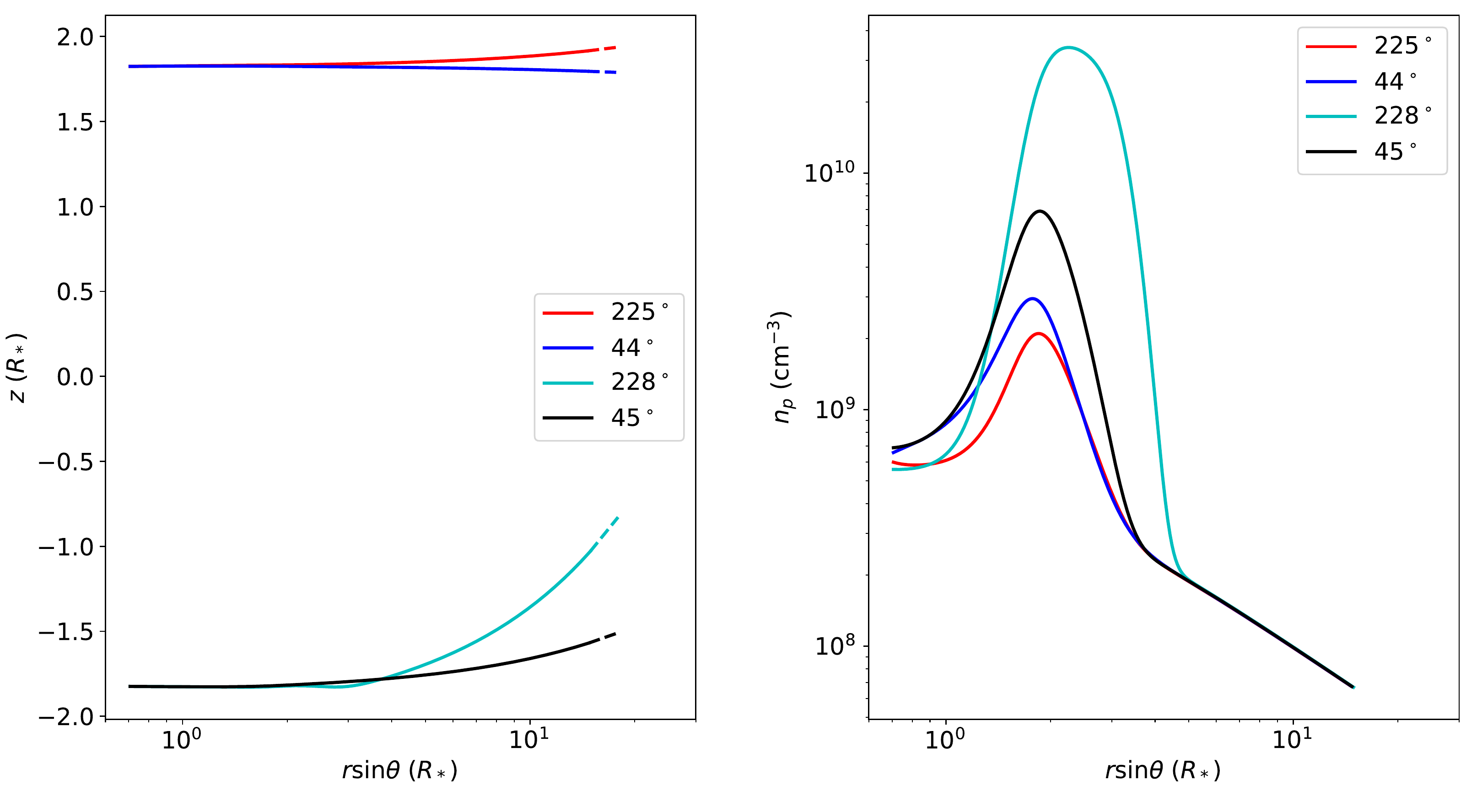}
\caption{\textit{Left:} The ray paths corresponding to different range of magnetic azimuthal coordinates $\phi_B$ that contribute to the observable pulses at 3 GHz (X-mode). The legends show the values of $\phi_B$ of the contributing auroral points. \textit{Right:} The number density experienced along the ray paths inside the IM.  \label{fig:ray_path_ne_asym_IM_3GHz}}
\end{figure*}

Finally the most important way in which the  refraction in the IM can affect the ECME lightcurve is to alter the sequence of arrival of pulses around the magnetic nulls. As mentioned  in the introduction (\S\ref{sec:intro}), we expect that when the north magnetic pole of the star is approaching and the south magnetic pole is receding (i.e. \bz~is changing from negative to positive, the rotational phases around null 1), the pulse from the south magnetic hemisphere will arrive before the pulse from the north magnetic hemisphere, and vice-versa. This expectation is based on the assumption that radiation produced at the north magnetic hemisphere always gets deviated upward and those at {\color{black} the} south magnetic hemispheres, gets deviated downward \citep[e.g. see Figure 2 of][]{leto2016}. 
While this assumption  holds good for the density profile considered in \S\ref{subsec:azimuth_sym}, or, for a constant density medium like the one considered in the `single refraction model', but need not be valid for the density profile given by Eq. \ref{eq:complex_asym_IM}. For example, we show paths followed by some of the rays that contribute to the observable pulses at 3 GHz in the left panel of Figure \ref{fig:ray_path_ne_asym_IM_3GHz}, and the corresponding densities along the ray path in the right panel. We find that the radiation from the south magnetic hemisphere undergo significant deviation upward (instead of downward). For the case of north magnetic hemisphere, some of the rays get deviated slightly upward and the rest deviate downward. The net result is that we see opposite sequence of arrival of pulses near the magnetic nulls.

In reality, we often see ECME pulses not behaving ideally in terms of pulse shapes, their phases of arrival etc.. For example, \citet{das2019a} observed a double-peaked ECME pulse from HD\,142990 for which there is not any satisfactory explanation yet. For such cases, we can use a realistic model for the magnetosphere and try to find out whether any such non-ideality can be obtained from the propagation effect. This, in turn, will allow us to get a better idea about the density structure in the stellar magnetosphere.

%The case presented in this subsection displays one of the strengths of our framework, which is to use a realistic model for magnetosphere and try to find out whether any such non-ideality can be obtained from the propagation effect. This, in turn, will allow us to get a better idea about the density structure in the stellar magnetosphere.

\section{Discussion}\label{sec:summary}
{\color{black} The} ECME observed from hot magnetic stars can become a highly useful tool to probe their magnetospheres. Since different frequencies arise at different heights, they pass through different parts of the IM and can undergo significantly different deviations depending on the density distribution they encounter in the IM. Such a situation is more likely for a star with high obliquity. In that case, if the type of density distribution is known (e.g. from the RRM model), we can use the framework presented here to simulate multi-frequency lightcurves. By comparing these lightcurves with the observed ones, one can obtain the best-fit values for different parameters associated with the density distribution. 
{\color{black} While doing so, one must discard the assumption of density discontinuities at the IM boundary, and work with a density grid in stead which can be easily incorporated in the current code.}
This framework can also be useful in disentangling effects produced by a complex magnetic field and those due to propagation effect. Till now, various non-ideal properties observed in {\color{black} the} ECME lightcurves, such as offset from the magnetic nulls, absence of pulse at a particular circular polarization etc., have been attributed to the presence of a magnetic field more complex than a dipole \citep[e.g.][]{trigilio2000,das2019a,leto2019}. Here we show that even for an ideal dipolar magnetic field, the ECME lightcurve can be highly non-intuitive. To disentangle these effects from those produced by a complex magnetic field, multi-frequency observations will be instrumental.

%By observing ECME from a star over a wide range of frequencies, it is possible to extract quantitative information about the density distribution in the IM with the framework presented here, provided there is prior idea about 

%Till now, no quantitative model existed to study the propagation effect undergone by the ECME in the IM of the host star. The past works on propagation effect have considered single refraction at the boundary between middle and inner magnetosphere at the time of entering the latter \citep{trigilio2011,lo2012}. This work is meant to be a tool for the quantitative study of the propagation effect. We present a general framework to calculate ECME ray paths through the IM with an arbitrary density distribution. We also demonstrate how we can use this framework to calculate `lag's (the difference in pulse arrival times at two different frequencies) for different frequency pairs, which can then be compared with the observations. At the end, we show that a star with an azimuthally asymmetric density distribution in the inner magnetosphere (a result of high obliquity), can produce a highly non-intuitive ECME lightcurve.

Although this framework is general enough to work for a magnetic field more complex than a dipole, we have currently implemented only the dipole case in our code. We plan to generalize it in future. We emphasize that this framework will enable the scientific community to understand the `non-ideal' properties of ECME, and in the process of doing it, will let us probe the density structure in the magnetosphere of the host star.

\section*{Acknowledgement}
BD thanks Gregg A. Wade (Royal Military College of Canada) for the discussions on the work presented here. BD thanks Gregg Hallinan (Caltech, USA) for pointing out the inadequacies of the earlier version of the code which helped her in improving the framework. BD thanks Veronique Petit (University of Delaware, USA) for providing all the visualizations obtained from the RRM model and also the motivation to go from 2D to 3D. PC acknowledges support from the Department of Science and Technology via SwarnaJayanti Fellowship awards (DST/SJF/PSA-01/2014-15). We acknowledge support of the Department of Atomic Energy, Government of India, under project no. 12-R\&D-TFR-5.02-0700.

\bibliography{das}

\appendix
{\color{black}
\section{Formulae for refractive indices}
The refractive index $\mu$, for the X and O modes are given by \citep{lee2013}:
\begin{align}
\mu_X^2=1-\frac{\nu_p^2}{\nu(\nu+\tau\nu_\mathrm{B})},\qquad
\mu_O^2=1-\frac{\tau\nu_p^2}{\nu(\tau\nu-\nu_\mathrm{B}\cos^2\psi)}\label{eq:refractive_index}
\end{align}
Here $\psi$ is the angle between the wave vector and the local magnetic field, $\nu_p$ and $\nu_\mathrm{B}$ are the local plasma frequency and the electron cyclotron frequency respectively, and $\tau$ is given by the following equation:

\begin{align*}
\tau=(\sigma+\sqrt{\sigma^2+\cos^2\psi})\frac{\nu_p^2-\nu^2}{|\nu_p^2-\nu^2|},\qquad
\sigma=\frac{\nu\nu_\mathrm{B}\sin^2\psi}{2|\nu^2-\nu_p^2|}
\end{align*}
}
\section{Rotation matrices for going from the line of sight frame of reference to that of the magnetic field}\label{sec:rot_matrix}
\begin{figure}
\centering
\includegraphics[trim={0cm 2cm 5cm 2cm},clip,width=0.45\textwidth]{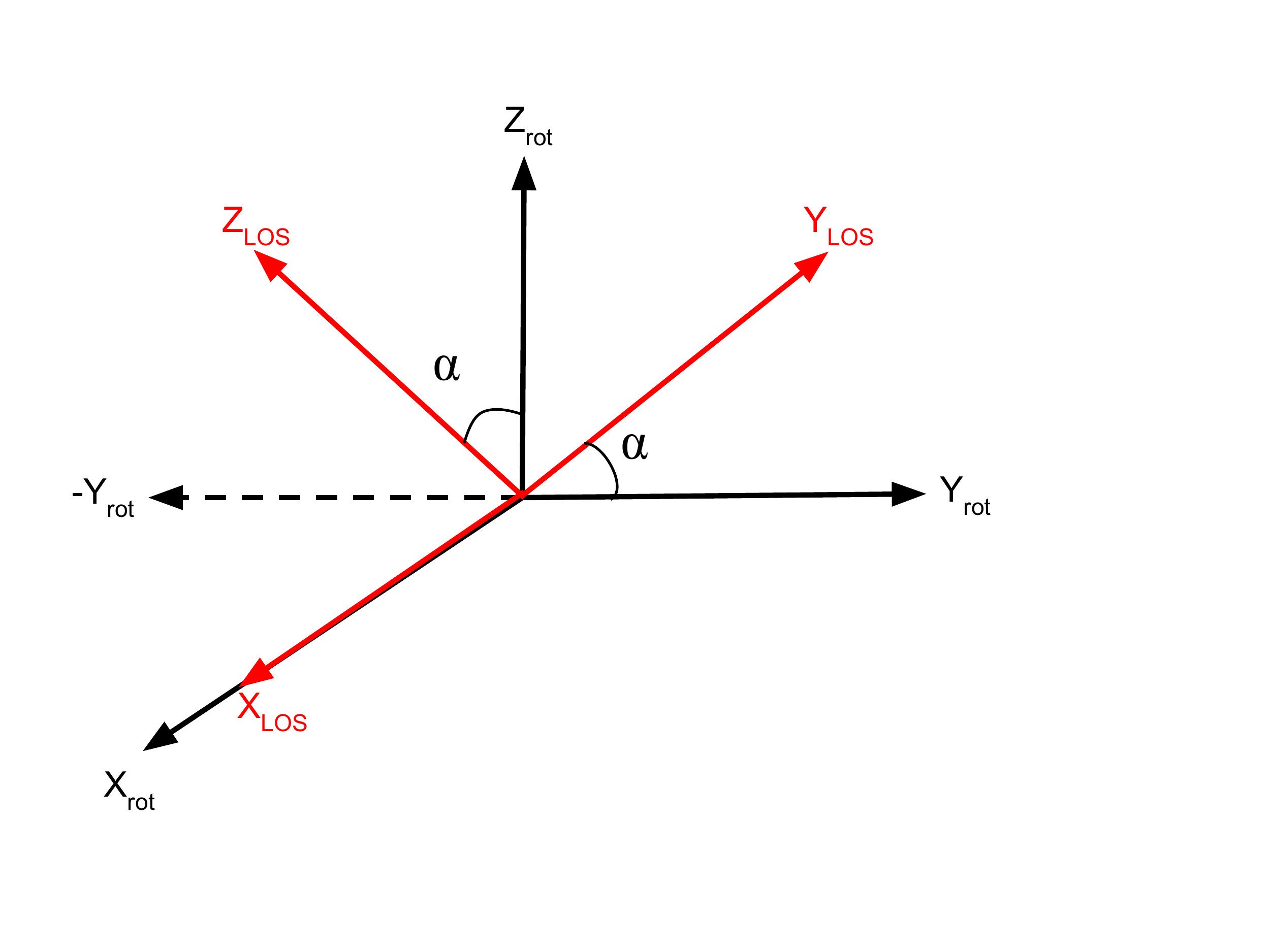}
\caption{The relation between the LoS and rotational frame of reference. The $X$ axes of both frames are aligned. The $Z$ axis of the rotation frame is along the rotation axis and that in case of the LoS frame is along the LoS. The LoS lies in the $-YZ$ plane of the rotational frame and is inclined by an angle $\alpha$ w.r.t. the rotation axis. The rotation matrix is given by Eq. \ref{eq:los_rot_frame}.\label{fig:los_rot_frame}}
\end{figure}

In order to obtain the lightcurve, we will have to define the three frames of reference, namely: the reference frame of the magnetic field, the reference frame of the rotation axis and the reference frame of the line of sight (LoS). In the magnetic frame of reference, we have already taken the $Z$ axis to lie along the dipole axis; in the rotation frame, the $Z$ axis lies along the rotation axis, and in the LoS frame, the $Z$ axis lies along the LoS. We first write the rotation matrix for getting the components of a vector $\textbf{X}$ in the star's rotation frame from the known components in the LoS frame. Let the inclination angle be $\alpha$ and the obliquity be $\beta$. we choose the LoS to lie in the $YZ$ plane of the rotation frame as shown in Figure \ref{fig:los_rot_frame}.

From Figure \ref{fig:los_rot_frame}, we get:
\begin{align}
\textbf{X}_\mathrm{rot}=
\begin{pmatrix}
  1 & 0 & 0\\ 
  0 & \cos\alpha & -\sin\alpha \\
  0 & \sin\alpha & \cos\alpha 
\end{pmatrix}
\textbf{X}_\mathrm{LoS} \label{eq:los_rot_frame}
\end{align}

\begin{figure}
\centering
\includegraphics[trim={0cm 2cm 5cm 2cm},clip,width=0.45\textwidth]{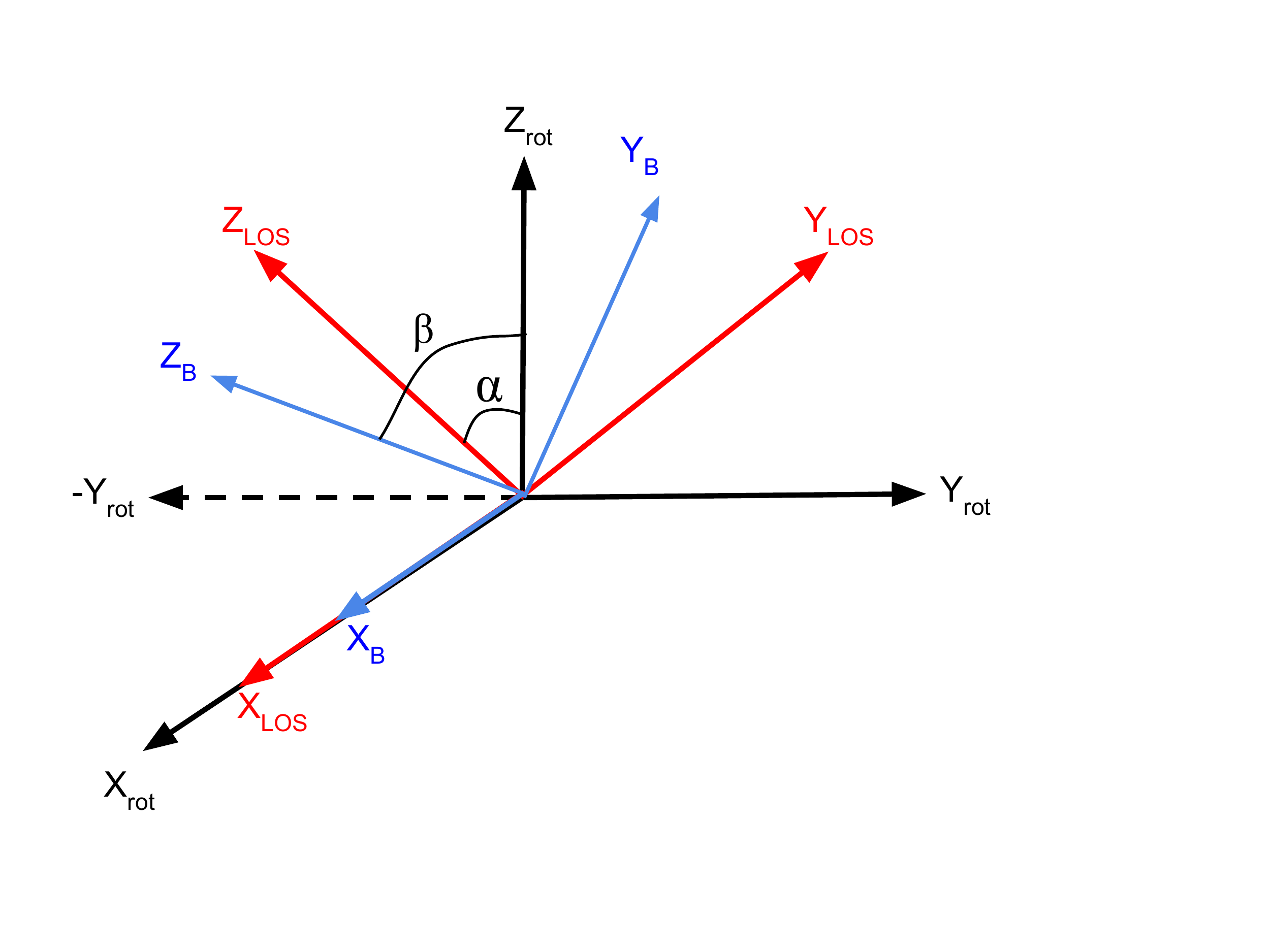}
\caption{The relation among the LoS, rotation and magnetic frames of reference when the rotational phase is zero. At this phase, the LoS, rotation axis and the magnetic dipole axes lie in the same plane and the $X$ axes of respective frames of reference are aligned. The magnetic axis makes an angle of $\beta$ w.r.t. the rotation axis and the angle between the LoS and the magnetic dipoel axis is $|\beta-\alpha|$ at this rotational phase. \label{fig:los_rot_B_frame_zero_rot_phase}}
\end{figure}

\begin{figure*}
\centering
\includegraphics[trim={0cm 2cm 5cm 2cm},clip,width=0.45\textwidth]{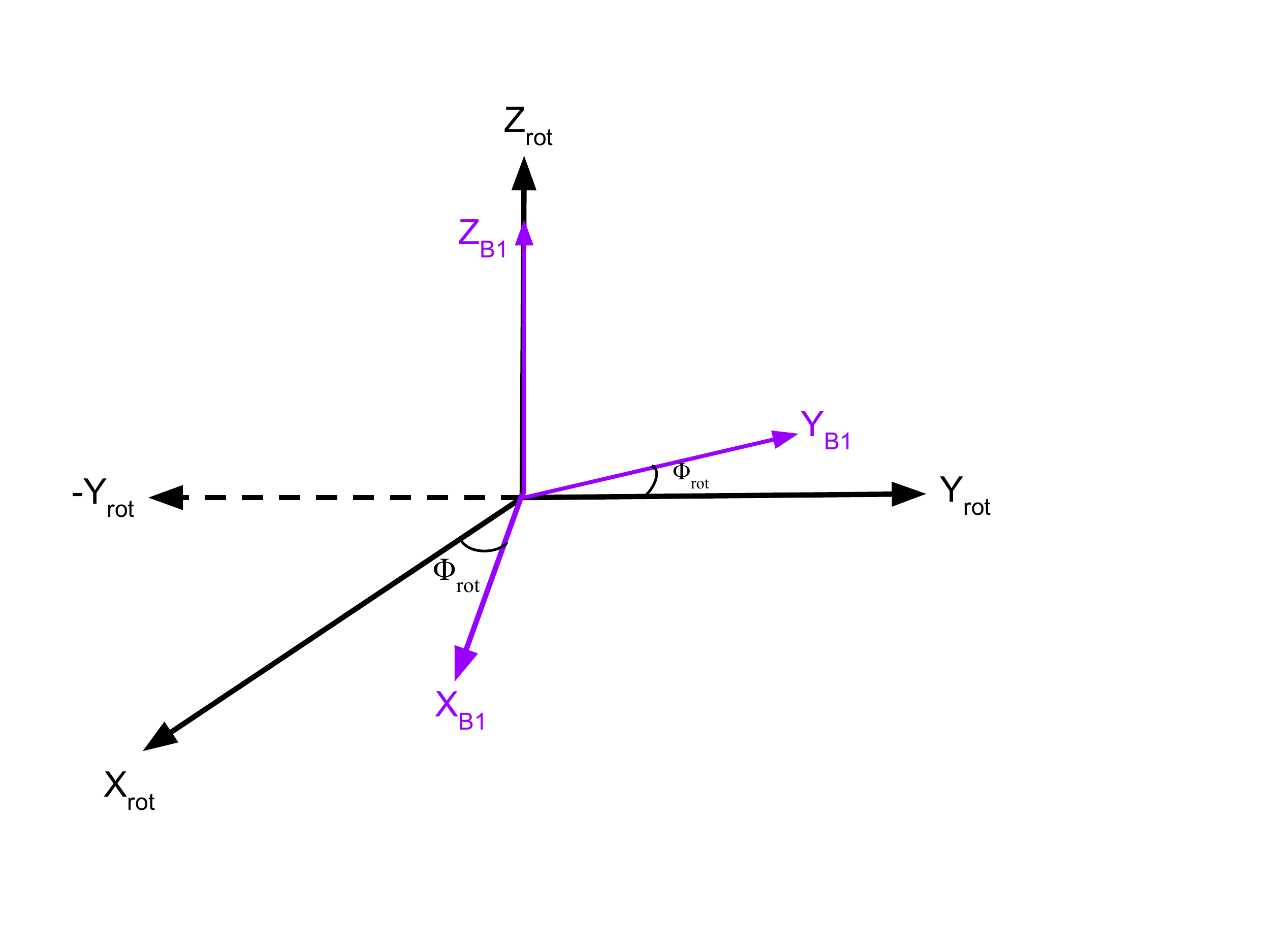}
\includegraphics[trim={0cm 2cm 5cm 2cm},clip,width=0.45\textwidth]{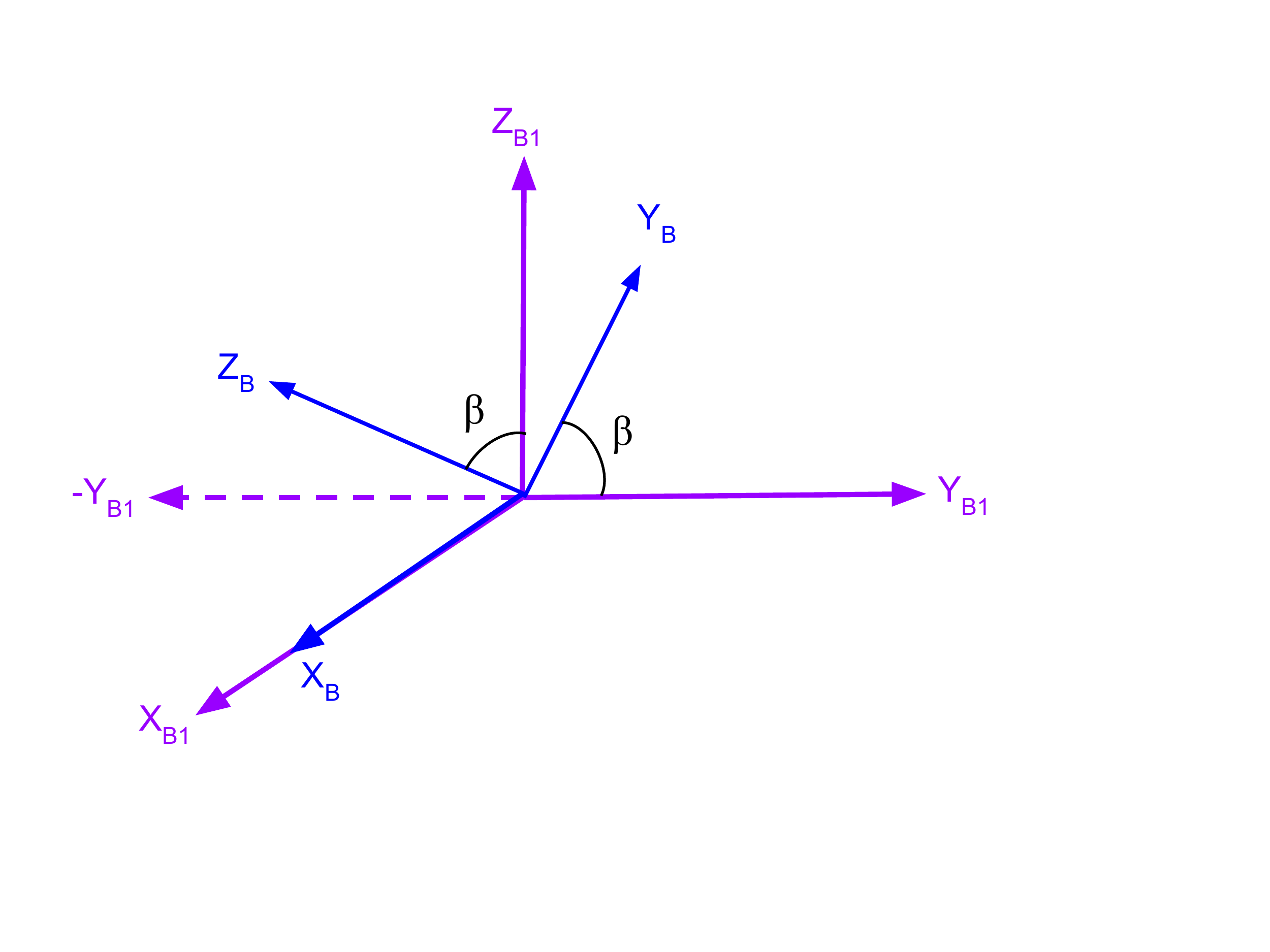}
\caption{The relation between magnetic and rotation frame for a given rotational  phase $\phi_\mathrm{rot}/2\pi$. We can get this relation by first considering the figure on the left panel, where we have the the $Z$ axes of frames are aligned for both frames ($B_1$ frame, $\beta=0$, and the other two are inclined by an angle of $\phi_\mathrm{rot}$. We next consider the relation between this intermediate $B_1$ frame and the magnetic frame (right panel), where now the $X$ axes are aligned and the other two are inclined by an angle $\beta$. The rotation matrix is given by Eq. \ref{eq:B_rot_frame}\label{fig:B_rot_frame}}
\end{figure*}

The above relation allows us to go from the LoS frame of reference to the star's rotation frame of reference. Next we have to find the rotation matrix for going from the rotation to the magnetic frame of reference. We define the zero of the rotational phases in such a way that the LoS, rotation axis and the magnetic axis lie in the same plane (Figure \ref{fig:los_rot_B_frame_zero_rot_phase}). We will find the rotation matrix between the magnetic and rotation frame for any rotational phase in two steps. In the first step, we take $\beta=0$ and the other two axes in one frame are misaligned with their counterparts in the other frame by an angle $\phi_\mathrm{rot}$ (rotational phase is $\phi_\mathrm{rot}/2\pi$). This is shown in the left panel of Figure \ref{fig:B_rot_frame}. We denote the axes in this `intermediate magnetic frame' as $X_\mathrm{B1}$, $Y_\mathrm{B1}$ and $Z_\mathrm{B1}$.

From the left panel of Figure \ref{fig:B_rot_frame}, we get:
\[\textbf{X}_\mathrm{B1}=
\begin{pmatrix}
  \cos\phi_\mathrm{rot} & \sin\phi_\mathrm{rot} & 0\\ 
  -\sin\phi_\mathrm{rot} & \cos\phi_\mathrm{rot} & 0 \\
  0 & 0 & 1
\end{pmatrix}
\textbf{X}_\mathrm{rot}
\]

In the second step, we consider the rotation matrix between this intermediate frame and the real magnetic frame. From the right panel of Figure \ref{fig:B_rot_frame}, we have:

\begin{align}
\textbf{X}_\mathrm{B}&=
\begin{pmatrix}
 1 & 0 & 0\\ 
  0 & \cos\beta & \sin\beta \\
  0 & -\sin\beta & \cos\beta
\end{pmatrix}
\textbf{X}_\mathrm{B1} \nonumber\\
&=
\begin{pmatrix}
\cos\phi_\mathrm{rot} & \sin\phi_\mathrm{rot} & 0\\
-\cos\beta\sin\phi_\mathrm{rot} &\cos\beta\cos\phi_\mathrm{rot} & \sin\beta \\
\sin\beta\sin\phi_\mathrm{rot} & -\sin\beta\cos\phi_\mathrm{rot} & \cos\beta 
\end{pmatrix}
\textbf{X}_\mathrm{rot}\label{eq:B_rot_frame}
\end{align}

Substituting from Eq. \ref{eq:los_rot_frame} in Eq. \ref{eq:B_rot_frame}, we get the rotation matrix for going from the LoS to the magnetic frame of reference, which is:
\begin{align}
\begin{pmatrix}
\cos\phi_\mathrm{rot} & \cos\alpha\sin\phi_\mathrm{rot} & -\sin\alpha\sin\phi_\mathrm{rot}\\
-\cos\beta\sin\phi_\mathrm{rot} & \cos\alpha\cos\beta\cos\phi_\mathrm{rot} & -\sin\alpha\cos\beta\cos\phi_\mathrm{rot}\\
& +\sin\alpha\sin\beta & +\cos\alpha\sin\beta\\
\sin\beta\sin\phi_\mathrm{rot} & -\cos\alpha\sin\beta\cos\phi_\mathrm{rot} & \sin\alpha\sin\beta\cos\phi_\mathrm{rot}\\
 & +\sin\alpha\cos\beta & +\cos\alpha\cos\beta
\end{pmatrix} \label{eq:B_los_frame}
\end{align}

Using this matrix, we can write the line of sight vector in the magnetic frame of reference ($\textbf{LoS}_\mathrm{B}$) for a given rotational phase $\phi_\mathrm{rot}/2\pi$. %The rest of the procedure is described in \S\ref{subsec:lightcurve}.

\section{Lightcurves of ECME for a given density profile in the IM}\label{sec:ecme_lightcurve}
The lightcurves at a given frequency can be obtained by the following steps:

\begin{enumerate}
\item Obtain $\textbf{k}_\mathrm{out}$ vectors for the full auroral circles (near both magnetic hemispheres). The auroral circles are defined by the frequency of ECME, harmonic number and the magnetic field strength (\S\ref{sec:framework}).
\item For each rotational phase, obtain the components of the line of sight (LoS) vector in the magnetic frame of reference (Eq. \ref{eq:B_los_frame}).
\item For that rotational phase, calculate the angle $\theta_k$, which is the angle between the line of sight vector and a given $\textbf{k}_\mathrm{out}$.
\item Add a contribution of $\exp\left(-{\theta_k}^2/{\sigma_\theta}^2\right)$ to the lightcurve for that rotational phase, where $\sigma_\theta$ is a measure of the ECME beam width.
\item Repeat step 2 to step 4 for each rotational phase.
\end{enumerate}

\end{document}